\documentclass[
	twocolumn,
	onecolappendix,
]{aastex701}

\usepackage{
  graphicx,
  bm,
  amsmath, 
  amssymb,
  mathtools,
  tikz,
  mathdots,
  tabularx,
  hyperref,
  xcolor,
}

\usepackage[frozencache=true,cachedir=minted-cache]{minted}

\newcommand{\entity}{\texttt{Entity}}

\newcommand{\nPhalf}{n+1/2}
\newcommand{\iPhalf}{i+1/2}
\newcommand{\jPhalf}{j+1/2}
\newcommand{\kPhalf}{k+1/2}

\newcommand{\AtN}[2]{{}^{(\mathtt{#2})}{#1}}
\newcommand{\AtNIJK}[5]{{}^{(\mathtt{#2})}{#1_{(\mathtt{#3,#4,#5})}}}
\newcommand{\CovAtNIJK}[6]{{}^{(\mathtt{#3})}{#1_{#2(\mathtt{#4,#5,#6})}}}

\newcommand{\AtNIJKtwoD}[4]{{}^{(\mathtt{#2})}{#1_{(\mathtt{#3,#4})}}}
\newcommand{\CovAtNIJKtwoD}[5]{{}^{(\mathtt{#3})}{#1_{#2(\mathtt{#4,#5})}}}

\setminted{fontsize=\small} 
\graphicspath{{./}{figures/}}

\begin{document}

\title{\entity~-- Hardware-agnostic Particle-in-Cell Code for Plasma Astrophysics.\\ I: Curvilinear Special Relativistic Module}


\correspondingauthor{Hayk Hakobyan}

\author[orcid=0000-0001-8939-6862]{Hayk Hakobyan}
\affiliation{Center for Computational Astrophysics, Flatiron Institute, New York, NY 10010, USA}
\affiliation{Physics Department \& Columbia Astrophysics Laboratory, Columbia University, New York, NY 10027, USA}
\affiliation{Computational Sciences Department, Princeton Plasma Physics Laboratory (PPPL), Princeton, NJ 08540, USA}
\email[show]{haykh.astro@gmail.com}

\author[orcid=0000-0003-4690-2774]{Ludwig M. B\"oss}
\affiliation{Department of Astronomy and Astrophysics, The University of Chicago, William Eckhart Research Center, Chicago, IL 60637}
\email{...}

\author[orcid=0000-0002-9074-4657]{Yangyang Cai}
\affiliation{Tsung-Dao Lee Institute of Shanghai Jiao Tong University, Pudong New Area, Shanghai, 201210, China}
\email{...}

\author[orcid=0000-0001-5121-1594]{Alexander Chernoglazov}
\affiliation{Institute for Advanced Study, Princeton, NJ 08540, USA }
\email{...}

\author[orcid=0000-0002-6300-3191]{Alisa Galishnikova}
\affiliation{Center for Computational Astrophysics, Flatiron Institute, New York, NY 10010, USA}
\email{...}

\author[orcid=0000-0001-9073-8591]{Evgeny A.  Gorbunov}
\affiliation{Department of Physics, University of Maryland, College Park, MD 20742, USA}
\email{...}

\author[orcid=0000-0002-5349-7116]{Jens F. Mahlmann}
\affiliation{Department of Physics and Astronomy, Dartmouth College, Hanover, NH 03755, USA}
\email{...}

\author[orcid=0000-0001-7801-0362]{Alexander Philippov}
\affiliation{Department of Physics, University of Maryland, College Park, MD 20742, USA}
\affiliation{Institute for Research in Electronics and Applied Physics, University of Maryland, College Park, MD 20742, USA}
\email{...}

\author[orcid=0000-0001-6541-734X]{Siddhant Solanki}
\affiliation{Department of Physics, University of Maryland, College Park, MD 20742, USA}
\affiliation{Institute for Research in Electronics and Applied Physics, University of Maryland, College Park, MD 20742, USA}
\email{...}

\author[orcid=0000-0002-3643-9205]{Arno Vanthieghem}
\affiliation{Sorbonne Universit\'e, Observatoire de Paris, Universit\'e PSL, CNRS, LUX, F-75005 Paris, France}
\email{...}

\author[orcid=0000-0003-0709-7848]{Muni Zhou}
\affiliation{Department of Physics and Astronomy, Dartmouth College, Hanover, NH 03755, USA}
\email{...}

\collaboration{all}{\entity~Development Team}

\begin{abstract}
  Entity is a new-generation, fully open-source particle-in-cell (PIC) code developed to overcome key limitations in astrophysical plasma modeling, particularly the extreme separation of scales and the performance challenges associated with evolving, GPU-centric computing infrastructures. It achieves hardware-agnostic performance portability across various GPU and CPU architectures using the Kokkos library. Crucially, Entity maintains a high standard for usability, clarity, and customizability, offering a robust and easy-to-use framework for developing new algorithms and grid geometries, which allows extensive control without requiring edits to the core source code. This paper details the core general-coordinate special-relativistic module. Entity is the first PIC code designed to solve the Vlasov-Maxwell system in general coordinates, enabling a coordinate-agnostic framework that provides the foundational structure for straightforward extension to arbitrary coordinate geometries. The core methodology achieves numerical stability by solving particle equations of motion in the global orthonormal Cartesian basis, despite using generalized coordinates like Cartesian, axisymmetric spherical, and quasi-spherical grids. Charge conservation is ensured via a specialized current deposition technique using conformal currents. The code exhibits robust scalability and performance portability on major GPU platforms (AMD MI250X, NVIDIA A100, and Intel Max Series), with the 3D particle pusher and the current deposition operating efficiently at about 2 nanoseconds per particle per timestep. Functionality is validated through a comprehensive suite of standard Cartesian plasma tests and the accurate modeling of relativistic magnetospheres in curvilinear axisymmetric geometries.
\end{abstract}


\section{Introduction}
\label{sec:intro}

Particle-in-cell (PIC) algorithms \citep{Dawson_1983, Hockney.Eastwood_1988, Birdsall.Langdon_1991, Villasenor.Buneman_1992} have been used extensively since the end of the last century to study plasma phenomena ranging from laboratory plasma applications \citep{Verboncoeur_2005, Esarey.etal_2009, Macchi.etal_2013}, material science \citep{Ding.etal_2020, Ngirmang.etal_2025}, space physics \citep{Lapenta_2012, Li.etal_2021}, and, ultimately, astrophysics \citep{Zenitani.Hoshino_2005, Spitkovsky2008a, Sironi.Spitkovsky_2014, Guo2014, Zhdankin_2017, Comisso_2019}. Particularly in the past two decades, the astrophysical community has expressed a growing interest in PIC, due to its ability to reconstruct particle distribution functions and make clear predictions on observational appearances of astrophysical objects and environments. Despite the growing complexities and capabilities of modern PIC codes, there are clear bottlenecks in the applicability of the algorithm in astrophysics, especially when modeling large-scale dynamics. Firstly, in astrophysics, the separation of global scales of the systems to plasma-kinetic scales is often extremely large, and resolving both dimensions is -- most of the time -- infeasible. This bottleneck is somewhat mitigated in some of the codes, which introduce curvilinear (so-called ``body-fitted'') grids that enhance the resolution in places where it is most necessary, while sacrificing it at large scales. Secondly, modern computing infrastructures are quickly moving towards more GPU-centric architectures, whereas most of the existing codes are explicitly designed to run on CPUs. And finally, code usability has long been an important barrier-of-entry for many PIC applications, with the existing codebases being greatly outdated, non-modular, and difficult (oftentimes, impossible) to modify and customize to more specific needs.

In this work, we introduce the fully open-source particle-in-cell code written in  \texttt{C++} -- \entity\footnote{Documentation: \href{https://entity-toolkit.github.io}{entity-toolkit.github.io}} -- designed to address all the issues raised above, while still maintaining a high standard for usability, clarity, and customizability. The goal of \entity~is to accommodate most of the use-cases of PIC in modern astrophysical research by implementing all of the features which have thus far been used by the community, while also providing a robust and easy-to-use framework for developing new algorithms. Currently, \entity~supports a range of different coordinate grids (with a simple capability of adding new ones), including general relativity (details will be published in paper II). \entity~also implements advanced field-solver stencils, and high-order particle shape functions (details will be published in paper III) which -- due to our coordinate-agnostic design -- work for an arbitrary grid geometry. We are also currently developing a set of modules for radiative, quantum-electrodynamic (QED), and hadronic processes, including two-body particle-productions and scatterings, the details for which will be published in future papers. Importantly, \entity~is built on an architecture-agnostic framework -- \texttt{Kokkos} -- which allows for easy performance portability on various GPU and CPU architectures.

In this paper, we introduce the basic philosophy behind the design of \entity, primarily focusing on the general-coordinate special-relativistic module. The paper is organized as follows. In Section~\ref{sec:methodology}, we discuss the PIC algorithm for general curvilinear coordinates. Section~\ref{sec:codestruct} introduces the general architectural structure of the code, describing the main data containers and archetypal kernels. In Section~\ref{sec:problem_generators} we discuss the main API -- the problem generators -- and describe how it enables the code's customizability and adaptability to various setups. Performance of the code on three major GPU architectures and its scalability are presented in Section~\ref{sec:performance}. Finally, in Section~\ref{sec:tests} we describe a set of standard problem generators included with the code, which help benchmark and test various features of the algorithm.

\section{Particle-in-cell Algorithm In Curvilinear Space}
\label{sec:methodology}

{\entity~} solves the Vlasov--Maxwell system of equations in flexible coordinate systems. In this section, we discuss numerical algorithms and implementations in special relativity. Currently, we support one-, two- and three-dimensional simulations in Cartesian coordinates and axisymmetric (2.5D) simulations in spherical coordinate systems, with applications to neutron star and black hole magnetospheres. We decided to avoid implementing 3D spherical geometry in favor of the so-called cubed-sphere numerical grid (currently under development), the details of which will be described in a future publication.


\subsection{Units and normalizations}
\label{sec:units}

In this section, we describe the \emph{physical} unit system we employ in \entity. Note, that internally, \entity~uses the so-called \emph{code units} (described in the next section). These are, ideally, invisible to the end-user, and all the conversions are -- most of the time -- performed implicitly. Most of the parameters defined in the configuration input file, or settings performed in the problem generator (described in section~\ref{sec:problem_generators}), as well as the output, are either in physical or dimensionless units.

Physical units in \entity~are based on Gaussian units, which we further dimensionlessize using multiple fiducial quantities. First of all, $c$ --- the speed of light in vacuum --- is set to unity, $c=1$. The user then has to provide the extent of the box, e.g., $(x_{\rm min}, x_{\rm max})$ in the 1D Cartesian case, or $(r_{\rm min}, r_{\rm max})$ in 2D spherical geometry. These distances are defined in \emph{physical units}, the interpretation of which is left to the user.\footnote{In spherical geometry, e.g., when modeling a magnetosphere of a neutron star, it is often advisable to use $r_{\rm min} = 1$, meaning that all the length-scales will be normalized to the radius of the star.} Time in these units is then naturally defined in terms of the light-crossing time of $1$ unit length: time to cross a 1D Cartesian box is then exactly $x_{\rm max} - x_{\rm min}$ units of time.

To further make the quantities dimensionless, we define the following fiducial quantities: $n_0$ -- fiducial number density, $B_0$ -- fiducial electric/magnetic field strength, $m_0$ -- fiducial mass, $4\pi q_0$ -- fiducial charge, and $V_0$ -- fiducial volume. More practically, these fiducial values are computed internally by the code using several user-provided quantities: $d_0 \equiv \left\{m_0 c^2 / (4\pi q_0^2 n_0)\right\}^{1/2}$ -- fiducial plasma skin-depth, $\rho_0 \equiv m_0 c^2 / (q_0 B_0)$ -- fiducial Larmor radius, $\mathtt{ppc_0} \equiv n_0 V_0$ -- fiducial number of particles per cell (PPC). The value of the fiducial volume, $V_0$, depends on the coordinate geometry, the resolution of the spatial grid, and the extent of the domain. For the regular Cartesian grid, it is simply $\Delta x^D$, where $\Delta x$ is the size of the cell, and $D$ is the dimensionality; for the 2D spherical geometry, it is inferred from the determinant of the metric, $\sqrt{\det{h_{ij}}}$, (discussed in the following section) at the corner of the grid: $r=r_{\rm min} + \Delta r/2$, and $\theta=\Delta \theta/2$.

Physical meanings of these quantities are as follows. Initializing $\mathtt{ppc_0}$ particles with weights $\left(\sqrt{\det{h_{ij}}}/V_0\right)^{1/D}$ per each cell of the simulation domain yields a uniform number density of $n_0$.\footnote{Note, that for Cartesian geometry, these weights will automatically be equal to $1$, since $\sqrt{\det h_{ij}}=V_0$.} $d_0$ is the skin-depth for relativistically cold pair-plasma of particles with masses $m_0$, charges $q_0$, and number density $n_0$. A particle of mass $m_0$ and charge $q_0$ with a four-velocity $|\bm{u}|=\gamma|\bm{\beta}| = 1$ moving perpendicular to a uniform magnetic field of strength $B_0$ has a Larmor radius of $\rho_0$. Using these fiducial values, we renormalize all of the quantities that the end-user interacts with, either in the setup of the simulation or in the output (except for the coordinates). Namely, all the particle number densities in the problem generator, as well as in the output, are normalized to $n_0$, the electric/magnetic fields -- to the fiducial value of $B_0$, while the electric currents are normalized to $4\pi q_0 n_0$.

Importantly, due to this choice of units, the physical results do not depend on either the spatial resolution of the box --- the number of cells --- or the sampling number of the distribution --- the number of particles per cell.


\subsection{Curvilinear coordinates in flat space-time}
\label{sec:curvspace}

In the special-relativistic module, \entity~solves all the equations in general curvilinear coordinates for a flat space-time (non-general-relativistic). We further consider only coordinates, for which the metric tensor is diagonal: $h_{ij}=0$ when $i\ne j$, which is valid for both the Cartesian and the 2D axisymmetric spherical coordinates considered in this paper.

Maxwell's equations in general coordinates can be written as ($c$ is set to $1$)

\begin{equation}
	\label{eq:curvspace-maxwell}
	\begin{aligned}
		\frac{\partial B^i}{\partial t} & =-\frac{1}{\sqrt{h}}\varepsilon^{ijk}\partial_j \underbrace{h_{km} E^m}_{E_k},           \\
		\frac{\partial E^i}{\partial t} & =\frac{1}{\sqrt{h}}\varepsilon^{ijk}\partial_j \underbrace{h_{km} B^m}_{B_k} - 4\pi J^i,
	\end{aligned}
\end{equation}

\noindent where $\partial_j\equiv \partial/\partial x^j$, $h\equiv \det{h_{ij}}$, $E^i$ and $B^i$ are the \textbf{contravariant} (upper index) components of the electric and magnetic fields, while $J^i$ is the electric current density.

Note that in general the coordinate basis $x^i$ does not have to be normalized, and thus the components of $E^i$ and $B^i$ are not straightforward to interpret physically. Because of that, we can use the forward and backward \textbf{tetrads}, $e^{i}_{\hat{j}}$ and $e^{\hat{i}}_{j}$, to transform an arbitrary vector $A^i$ to and from the orthonormal basis:

\begin{equation}
	\label{eq:curvspace-tetradtrans}
	A^i =e^{i}_{\hat{j}}A^{\hat{j}},~\text{and}~A^{\hat{i}} =e^{\hat{i}}_{j}A^j.
\end{equation}

\noindent For a diagonal metric tensor, the tetrad transformation matrix is also diagonal, and is equal to $e^{\hat{i}}_i=\left(e^i_{\hat{i}}\right)^{-1}=\sqrt{h_{ii}}$. While \entity~solves Maxwell's equations using the contravariant field components, it also uses the tetrad transformations to convert the fields to/from the orthonormal basis: e.g., when mapping the user input (which is done in the physical coordinates), when outputting the fields, or when pushing the particles (discussed in a later section).

Equations of motion for each individual macro-particle $p$ of species $s$ with positions, $x_p^i$ (contravariant), and four-velocities, $u_{p,i}$ (\textbf{covariant}, or lower index), written in general curvilinear coordinates have additional curvature terms. To make their numerical integrations easier, we instead solve these in the global orthonormal Cartesian basis (denoted further with $\hat{X}$): $\bm{x}_p^{\hat{X}}$, $\bm{u}_p^{\hat{X}}$:

\begin{equation}
	\label{eq:curvspace-equations-of-motion}
	\begin{aligned}
		\frac{d\bm{u}_p^{\hat{X}}}{dt} & =\frac{q_s}{m_s }\left[\bm{E}^{\hat{X}}_p+\frac{\bm{u}_p^{\hat{X}}}{\gamma_p}\times \bm{B}^{\hat{X}}_p\right], \\
		\frac{d\bm{x}_p^{\hat{X}}}{dt} & =\frac{\bm{u}_p^{\hat{X}}}{\gamma_p},                                                                          \\
		\textrm{where}~\gamma_p        & \equiv\sqrt{1+|\bm{u}_p^{\hat{X}}|^2}.
	\end{aligned}
\end{equation}

\noindent With this choice, the system \eqref{eq:curvspace-equations-of-motion} can be solved using conventional particle pusher algorithms, \citep[e.g.,][]{Boris_1970, Vay_2008}, with the added complexity of transforming back and forth from the general curvilinear basis to the global Cartesian one.

Charge conservation in curvilinear coordinates can be written similarly to the regular Cartesian case by applying $\partial_i$ to Amp\'ere's law in \eqref{eq:curvspace-maxwell}:

\begin{equation}
	\label{eq:curvspace-chargecons-1}
	\frac{\partial}{\partial t}\partial_i \left(\sqrt{h} E^i\right) = -4\pi \partial_i \left(\sqrt{h} J^i\right),
\end{equation}

\noindent We can then directly enforce Poisson's law, which in curvilinear coordinates is written as:

\begin{equation}
	\frac{1}{\sqrt{h}}\partial_i\left(\sqrt{h}E^i\right)=4\pi \rho,
\end{equation}

\noindent with $\rho$ being the charge density. As we discuss further, by employing a coordinate-conformal particle shape function, $S(\bm{x}-\bm{x}_p)$, where $\bm{x}_p$ is the position of particle $p$ in general curvilinear coordinate basis, $\bm{x}$, we can express the charge density as: $\rho = \sum_s q_s\sum_p (1/\sqrt{h})S(\bm{x}-\bm{x}_p)$, where the index $s$ corresponds to individual particle species ($q_s$ is the charge for each species), while the index $p$ is for each individual particle. Then the charge conservation from~\eqref{eq:curvspace-chargecons-1} can be written as:

\begin{equation}
	\label{eq:curvspace-chargecons-2}
	\sum_s q_s\sum_p\frac{\partial}{\partial t}S(\bm{x}-\bm{x}_p)+\partial_i\mathcal{J}^i=0,
\end{equation}

\noindent where we denote $\mathcal{J}^i\equiv \sqrt{h}J^i$, and will refer to this quantity as the \textbf{conformal currents}. Equation \eqref{eq:curvspace-chargecons-2} is directly analogous to the Cartesian charge conservation law, with the exception that the current density used is now weighted by $\sqrt{h}$. Thus, any charge-conservative current deposition algorithm designed to solve in the normal Cartesian case \citep[e.g.,][]{Villasenor.Buneman_1992, Esirkepov_2001, Umeda.etal_2003}, can be directly applied to solve it in arbitrary curvilinear coordinates, with the additional step of recovering the actual current densities, $J^i=\mathcal{J}^i/\sqrt{h}$, before using them in Amp\'ere's law in \eqref{eq:curvspace-maxwell}.



\subsection{Time discretization \& PIC loop}
\label{sec:time-discretization}

Equations \eqref{eq:curvspace-maxwell}, \eqref{eq:curvspace-equations-of-motion}, and \eqref{eq:curvspace-chargecons-2}, fully determine the closed system to solve the Vlasov-Maxwell equations with the finitely sampled distribution function for species $s$ in general curvilinear basis: $f_s(t,x^i,u_i)\equiv\sum_p S(x^i-x_p^i)\delta(u_i-u_{p,i})/\sqrt{h}$, where $S(x^i)$ is the conformal shape function: $\int S d^3x^i\equiv 1$, $\delta$ is a Dirac delta-function, and $h\equiv h(x^i)\equiv \det{h_{ij}}$.\footnote{Note, that by choosing $S$ to be conformal, we ensure: $\int f_s d \Gamma\equiv \int f_s \sqrt{h} d^3x^i d^3u^i = N_s$, where $N_s$ is the total number of particles of species $s$, while the covariant integration, $\int d\Gamma$, is performed over the whole phase space.}

\entity~solves this system using the FDTD leap-frog algorithm, by discretizing the electric and magnetic field components in both time and space, and correspondingly, discretizing the particle positions and four-velocities in time. In time, the electric field components, as well as the particle positions, are co-aligned at $ \AtN{t}{n}$, while the magnetic field components and the particle four-velocities are staggered backwards by a half-timestep, $\AtN{t}{n-1/2}$. The electric currents, computed from particle positions $\AtN{x^i_p}{n}$ and $\AtN{x^i_p}{n+1}$ are thus defined at $\AtN{J^i}{n+1/2}$. The duration of each timestep is fixed at $\Delta t\equiv \AtN{t}{n} - \AtN{t}{n-1}$, such that $ \AtN{t}{n}\equiv \texttt{n}\Delta t$. To avoid confusion, here and further we use typewriter font, e.g., $\texttt{i}$, $\texttt{n}$, to indicate quantities related to the discretized numerical grid (both in time and space).

Below, we outline the full particle-in-cell routine step-by-step. For brevity, we ignore the spatial discretization, which we discuss in more detail in the next section.

\begin{enumerate}
	\item Start at $\AtN{t}{n}$ with the following quantities stored:
	      \begin{equation*}
		      \begin{aligned}
			       & \AtN{B^i}{n-1/2},~\AtN{E^i}{n};                  \\
			       & \AtN{\bm{u}_p^{\hat{X}}}{n-1/2},~\AtN{x_p^i}{n}.
		      \end{aligned}
	      \end{equation*}

	\item Compute $\AtN{B^i}{n}$ using $\AtN{E^i}{n}$:
	      \begin{equation*}
		      \AtN{B^i}{n}=\AtN{B^i}{n-1/2}-\frac{\Delta t}{2\sqrt{h}}\varepsilon^{ijk}\partial_j h_{km} \AtN{E^m}{n}.
	      \end{equation*}

	\item For each particle $p$, interpolate the fields from the grid to particle position $x_p^i$:
	      \begin{equation*}
		      \AtN{\{E_p^i,B_p^i\}}{n}\equiv \int \AtN{\{E^i,B^i\}}{n} S(x^i-x_p^i) d^3 x^i.
	      \end{equation*}

	\item Convert the interpolated fields and the coordinates to the global Cartesian basis:
	      \begin{equation*}
		      \begin{aligned}
			      \AtN{\{E_p^i,B_p^i\}}{n} & \longrightarrow \AtN{\{\bm{E}_p^{\hat{X}},\bm{B}_p^{\hat{X}}\}}{n}; \\
			      \AtN{x_p^i}{n}           & \longrightarrow \AtN{\bm{x}_p^{\hat{X}}}{n}.
		      \end{aligned}
	      \end{equation*}

	\item Update the four-velocities using $\AtN{\{\bm{E}_p^{\hat{X}},\bm{B}_p^{\hat{X}}\}}{n}$:
	      \begin{equation*}
		      \AtN{\bm{u}_p^{\hat{X}}}{n-1/2}\xRightarrow[\Delta t]{} \AtN{\bm{u}_p^{\hat{X}}}{n+1/2}.
	      \end{equation*}

	\item Update the position using $\AtN{\bm{u}_p^{\hat{X}}}{n+1/2}$:
	      \begin{equation*}
		      \AtN{\bm{x}_p^{\hat{X}}}{n}\xRightarrow[\Delta t]{} \AtN{\bm{x}_p^{\hat{X}}}{n+1}.
	      \end{equation*}

	\item Convert particle positions back to curvilinear basis:
	      \begin{equation*}
		      \AtN{\bm{x}_p^{\hat{X}}}{n+1}\longrightarrow \AtN{x_p^{i}}{n+1}.
	      \end{equation*}

	\item Deposit the conformal currents using $\AtN{x_p^{i}}{n}$ and $\AtN{x_p^{i}}{n+1}$:
	      \begin{equation*}
		      \begin{split}
			      \partial_i & \{\AtN{\mathcal{J}^i}{n+1/2}\} = -\sum_s q_s\cdot                             \\
			                 & \cdot \sum_p\frac{S(x^i-\AtN{x_p^i}{n+1}) - S(x^i-\AtN{x_p^i}{n})}{\Delta t}.
		      \end{split}
	      \end{equation*}

	\item Filter the conformal currents, $\AtN{\mathcal{J}^i}{n+1/2}$, using digital filter passes.

	\item Compute $\AtN{B^i}{n+1/2}$ using $\AtN{E^i}{n}$:
	      \begin{equation*}
		      \AtN{B^i}{n+1/2}=\AtN{B^i}{n}-\frac{\Delta t}{2\sqrt{h}}\varepsilon^{ijk}\partial_j h_{km} \AtN{E^m}{n}.
	      \end{equation*}

	\item Compute $\AtN{E^i}{n+1}$ using $\AtN{B^i}{n+1/2}$ and $\AtN{\mathcal{J}^i}{n+1/2}$:
	      \begin{equation*}
		      \begin{split}
			      \AtN{E^i}{n+1} = & \AtN{E^i}{n}+\frac{\Delta t}{\sqrt{h}}\varepsilon^{ijk}\partial_j h_{km} \AtN{B^m}{n+1/2} \\
			                       & -4\pi\Delta t\frac{\AtN{\mathcal{J}^i}{n+1/2}}{\sqrt{h}}.
		      \end{split}
	      \end{equation*}

	\item Finally at $\AtN{t}{n+1}$ we have the following quantities:
	      \begin{equation*}
		      \begin{aligned}
			       & \AtN{B^i}{n+1/2},~\AtN{E^i}{n+1};                  \\
			       & \AtN{\bm{u}_p^{\hat{X}}}{n+1/2},~\AtN{x_p^i}{n+1},
		      \end{aligned}
	      \end{equation*}
	      \noindent and the cycle can be repeated.
\end{enumerate}

This routine has the advantage of minimizing the number of arrays stored in memory simultaneously after each substep. In particular, we only store a single instance of each of the fields and particle four-velocities, and two instances of currents (for filtering) and particle coordinates: $\AtN{x_p^i}{n}$ and $\AtN{x_p^i}{n+1}$. Note also that particle four-velocities are always stored in the global Cartesian basis, $\AtN{\bm{u}_p^{\hat{X}}}{n+1/2}$, and are only converted to the curvilinear basis when the data is being output.


\subsection{Spatial discretization}
\label{sec:spatial-discretization}

Spatial discretization of $E^i$, $B^i$, and $J^i$ is done according to \cite{Yee_1966}, where the grid is equally spaced in general coordinate basis, and the nodes of the grid are positioned at $x^{i(\mathtt{i})}=\mathtt{i}\Delta x^i$ ($\Delta x^i$ is the cell size in the $i$-th direction). Thus, the grid itself is equally spaced in general curvilinear coordinates. The main difference with the more conventional Cartesian routines is the presence of the metric tensor. In \entity, these are simply given by a set of functions which compute $h_{ij}(x^i)$, and $\sqrt{h}(x^i)$ at specific locations; these values, however, are never stored in memory, but are, instead, computed per each component on-the-fly. The full set of discretized Maxwell's equations are presented in the appendix~\ref{app:maxwell-discrete}.

Up until now, we made no specific assumptions on the geometry of the grid, and the discretized Maxwell's equations \eqref{eq:discretized-faraday} and \eqref{eq:discretized-ampere} are applicable as long as the space-time is flat, and the metric tensor is diagonal. Further, it is useful to distinguish three special coordinate systems which are explicitly supported by the \entity:

\begin{itemize}
	\item 1D/2D/3D Cartesian coordinates:
	      \begin{equation*}
		      x^i\equiv \{x,~y,~z\};
	      \end{equation*}
	\item 2D axisymmetric spherical (polar) coordinates:
	      \begin{equation*}
		      x^i\equiv\{r,~\theta,~\phi\};
	      \end{equation*}
	\item 2D axisymmetric quasi-spherical coordinates:
	      \begin{equation*}
		      x^i\equiv\{\mathcal{R},~\mathcal{T},~\phi\};
	      \end{equation*}
	      where the quasi-spherical coordinates are defined through regular spherical ones via
	      \begin{equation*}
		      \begin{aligned}
			      \mathcal{R} & \equiv \log{(r-r_0)},                                                            \\
			      \theta      & \equiv\mathcal{T}+2\vartheta \mathcal{T}(1-2\mathcal{T}/\pi)(1-\mathcal{T}/\pi),
		      \end{aligned}
	      \end{equation*}
	      with $r_0$ and $\vartheta$ defining the radial/angular stretching of the spherical coordinates towards the origin/equator \citep[see, e.g.,][]{Porth.etal_2017}.
\end{itemize}

Notice that while in 2D axisymmetric geometry all the derivatives $\partial_\phi$ are, by construction, zero, and thus all the fields are effectively two-dimensional, particles still keep track of their $\phi$ position, which is used to transform back and forth the global Cartesian coordinate frame.

\begin{figure}
	\centering
	\begin{tikzpicture}[scale=1,>=latex]

		\def\nr{6}
		\def\nt{16}
		\def\rmin{3.0}
		\def\rmax{6.0}
		\def\rzero{2.0}
		\def\h{0.4}

		\pgfmathsetmacro{\Rmax}{ln(\rmax-\rzero)}
		\pgfmathsetmacro{\Rmin}{ln(\rmin-\rzero)}
		\pgfmathsetmacro{\dR}{(\Rmax-\Rmin)/\nr}

		\foreach \i in {0,...,\nr} {
				\pgfmathsetmacro{\r}{\rzero+exp(\Rmin+\dR*\i)}
				\draw[blue!50, line width=0.75, domain=90:-90, samples=100, variable=\t]
				plot ({\r*cos(\t)}, {\r*sin(\t)});
			}

		\foreach \j in {0,...,\nt} {
				\pgfmathsetmacro\angle{\j * 180 / \nt + 2 * \h * \j * 180 / \nt * (1 - 2 * \j / \nt)*(1 - \j / \nt)}

				\pgfmathsetmacro\rmx{\rmin * sin(\angle)}
				\pgfmathsetmacro\rmy{\rmin * cos(\angle)}
				\pgfmathsetmacro\rx{\rmax * sin(\angle)}
				\pgfmathsetmacro\ry{\rmax * cos(\angle)}
				\draw[red!50, line width=0.75] (\rmx,\rmy) -- (\rx,\ry);
			}

		\tikzset{
			rlabel/.style={
					font=\scriptsize,
					left,
					color=blue!65!white
				}
		}
		\pgfmathsetmacro\rphysmid{0.5*(\rmax+\rmin)}
		\node[left=12pt, color=blue!65!white] at (0,{\rzero+exp(\Rmin+\dR*\nr/2)}) {$\bm{\mathcal{R}}$};
		\foreach \y/\text in {
		{\rzero+exp(\Rmin)}/{$\mathcal{R}_{\rm min}$},
		{\rzero+exp(\Rmin+\dR)}/{$\mathcal{R}_{\rm min}+\Delta\mathcal{R}$},
		{\rzero+exp(\Rmin+\nr*\dR/2)}/{$\vdots$},
		{\rzero+exp(\Rmin+(\nr-1)*\dR)}/{$\mathcal{R}_{\rm max}-\Delta\mathcal{R}$},
		{\rzero+exp(\Rmin+\nr*\dR}/{$\mathcal{R}_{\rm max}$}
		} {
		\node[rlabel] at ({0}, {\y}) {\text};
		}

		\tikzset{
			x1label/.style={
					font=\scriptsize,
					left,
					color=blue!45!white
				}
		}
		\node[left=12pt, color=blue!45!white] at (0,{-\rzero-exp(\Rmin+\dR*\nr/2)}) {$\mathtt{x}^2$};
		\foreach \y/\text in {
		{-\rzero-exp(\Rmin)}/{$0$},
		{-\rzero-exp(\Rmin+\dR)}/{$1$},
		{-\rzero-exp(\Rmin+\dR*\nr/2)}/{$\vdots$},
		{-\rzero-exp(\Rmin+(\nr-1)*\dR)}/{$\mathtt{N}_{\mathcal{R}}-1$},
		{-\rzero-exp(\Rmin+\nr*\dR}/{$\mathtt{N}_{\mathcal{R}}$}
		} {
		\node[x1label] at ({0}, {\y}) {\text};
		}

		\tikzset{
			thlabelUp/.style={
					font=\scriptsize,
					below,
					color=red!65!white
				}
		}
		\tikzset{
			thlabelDwn/.style={
					font=\scriptsize,
					above,
					color=red!65!white
				}
		}
		\node[left=2pt, color=red!65!white] at (\rmin,0) {$\bm{\mathcal{T}}$};
		\foreach \theta/\text in {
		{0}/{$0$},
		{1 * 180 / \nt + 2 * \h * 1 * 180 / \nt * (1 - 2 * 1 / \nt)*(1 - 1 / \nt)}/{$\Delta\mathcal{T}$}
		} {
		\node[thlabelUp] at ({\rmin*sin(\theta)}, {\rmin*cos(\theta)}) {\text};
		}
		\foreach \theta/\text in {
		{(\nt - 1) * 180 / \nt + 2 * \h * (\nt - 1) * 180 / \nt * (1 - 2 * (\nt - 1) / \nt)*(1 - (\nt - 1) / \nt)}/{$\pi-\Delta\mathcal{T}$},
		{180}/{$\pi$}
		} {
		\node[thlabelDwn] at ({\rmin*sin(\theta)}, {\rmin*cos(\theta)}) {\text};
		}

		\tikzset{
			x2labelUp/.style={
					font=\scriptsize,
					above,
					color=red!45!white
				}
		}
		\tikzset{
			x2labelDwn/.style={
					font=\scriptsize,
					below,
					color=red!45!white
				}
		}
		\node[right=2pt, color=red!45!white] at (\rmax,0) {$\mathtt{x}^2$};
		\foreach \theta/\text in {
		{0}/{$0$},
		{1 * 180 / \nt + 2 * \h * 1 * 180 / \nt * (1 - 2 * 1 / \nt)*(1 - 1 / \nt)}/{$1$}
		} {
		\node[x2labelUp] at ({\rmax*sin(\theta)}, {\rmax*cos(\theta)}) {\text};
		}
		\foreach \theta/\text in {
		{(\nt - 1) * 180 / \nt + 2 * \h * (\nt - 1) * 180 / \nt * (1 - 2 * (\nt - 1) / \nt)*(1 - (\nt - 1) / \nt)}/{$\mathtt{N}_\mathcal{T}-1$},
		{180}/{$\mathtt{N}_\mathcal{T}$}
		} {
		\node[x2labelDwn] at ({\rmax*sin(\theta)}, {\rmax*cos(\theta)}) {\text};
		}

	\end{tikzpicture}
	\caption{Physical, $\mathcal{R}$-$\mathcal{T}$ ($r$-$\theta$), and code, $\mathtt{x}^1$-$\mathtt{x}^2$, coordinates of the 2D axisymmetric quasi-spherical grid. Here $r_0=(2/3)r_{\rm min}$, and $\vartheta = 0.4$.}
	\label{fig:spat-disc-qsphgrid}
\end{figure}
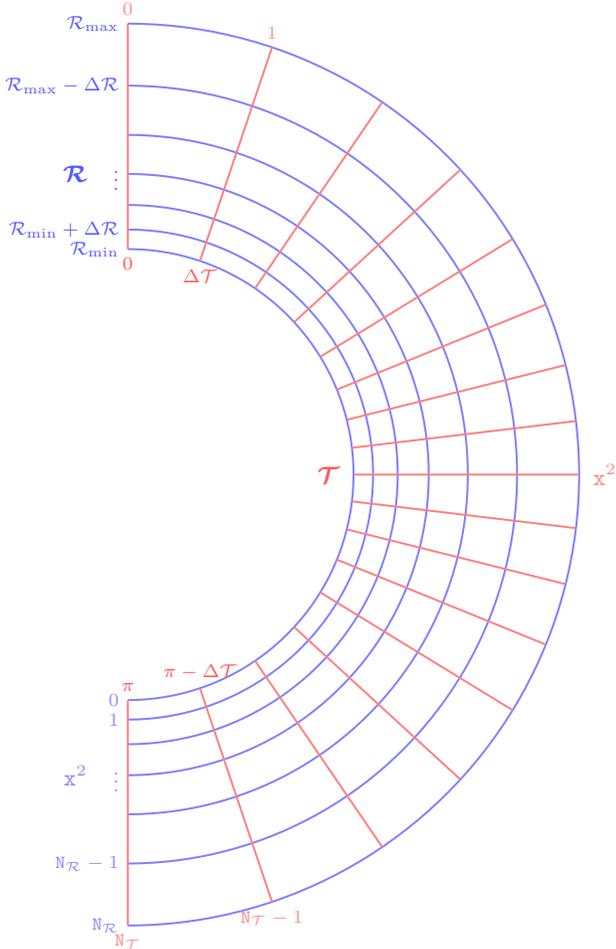

Internally, \entity~additionally remaps these physical coordinates, $x^i$, into the so-called \textbf{code-coordinates}, $\mathtt{x^i}$ using the number of grid cells in each dimension, $\mathtt{N_i}$:

\begin{equation}
	\mathtt{x^i}\equiv \mathtt{N_i}(x^i-x^i_{\rm min})/(x^i_{\rm max}-x^i_{\rm min}),
\end{equation}

\noindent where $x^i_{\rm min}$ and $x^i_{\rm max}$ correspond to the extent of the domain in physical units. The corresponding conventions for the 2D Cartesian, and 2D axisymmetric quasi-spherical grids are shown in Figures~\ref{fig:spat-disc-cartgrid} and~\ref{fig:spat-disc-qsphgrid}.

\begin{figure}
	\centering
	\begin{tikzpicture}[scale=1.0]

		\def\nx{8}
		\def\ny{4}
		\def\scalex{0.75}
		\def\scaley{0.75}

		\foreach \i in {0,...,\nx} {
				\draw[blue!50, line width=0.75] (\i*\scalex,0) -- (\i*\scalex,\ny*\scaley);
			}
		\foreach \j in {0,...,\ny} {
				\draw[red!50, line width=0.75] (0,\j*\scaley) -- (\nx*\scalex,\j*\scaley);
			}

		\tikzset{
			xlabel/.style={
					font=\scriptsize,
					below,
					color=blue!65!white
				}
		}
		\node[below=12pt, color=blue!65!white] at (\nx*\scalex/2,0) {$\bm{x}$};
		\foreach \x/\y/\text in {
		{0}/{-0.5}/{$x_{\rm min}$},
		{1}/{0}/{$x_{\rm min}+\Delta x$},
		{(\nx/2)}/{0}/{$\dots$},
		{(\nx-1)}/{0}/{$x_{\rm max}-\Delta x$},
		{\nx}/{-0.5}/{$x_{\rm max}$}
		} {
		\node[xlabel] at ({\x * \scalex}, {\y * \scaley}) {\text};
		}

		\tikzset{
			x1label/.style={
					font=\scriptsize,
					above,
					color=blue!45!white
				}
		}
		\node[above=12pt, color=blue!45!white] at (\nx*\scalex/2,\ny*\scaley) {$\mathtt{x}^1$};
		\foreach \x/\text in {
		{0}/{$0$},
		{1}/{$1$},
		{(\nx/2)}/{$\dots$},
		{(\nx-1)}/{$\mathtt{N_x}-1$},
		{\nx}/{$\mathtt{N_x}$}
		} {
		\node[x1label] at ({\x * \scalex}, {\ny * \scaley}) {\text};
		}

		\tikzset{
			ylabel/.style={
					font=\scriptsize,
					left,
					color=red!65!white
				}
		}
		\node[left=12pt, color=red!65!white] at (0,\ny*\scaley/2) {$\bm{y}$};
		\foreach \y/\text in {
		{0}/{$y_{\rm min}$},
		{1}/{$y_{\rm min}+\Delta y$},
		{(\ny/2)}/{$\vdots$},
		{(\ny-1)}/{$y_{\rm max}-\Delta y$},
		{\ny}/{$y_{\rm max}$}
		} {
		\node[ylabel] at ({0}, {\y * \scaley}) {\text};
		}

		\tikzset{
			x2label/.style={
					font=\scriptsize,
					right,
					color=red!45!white
				}
		}
		\node[right=12pt, color=red!45!white] at (\nx*\scalex,\ny*\scaley/2) {$\mathtt{x}^2$};
		\foreach \y/\text in {
		{0}/{$0$},
		{1}/{$1$},
		{(\ny/2)}/{$\vdots$},
		{(\ny-1)}/{$\mathtt{N_y}-1$},
		{\ny}/{$\mathtt{N_y}$}
		} {
		\node[x2label] at ({\nx * \scalex}, {\y * \scaley}) {\text};
		}

	\end{tikzpicture}
	\caption{Physical, $x$-$y$, and code, $\mathtt{x}^1$-$\mathtt{x}^2$, coordinates of the 2D Cartesian grid.}
	\label{fig:spat-disc-cartgrid}
\end{figure}
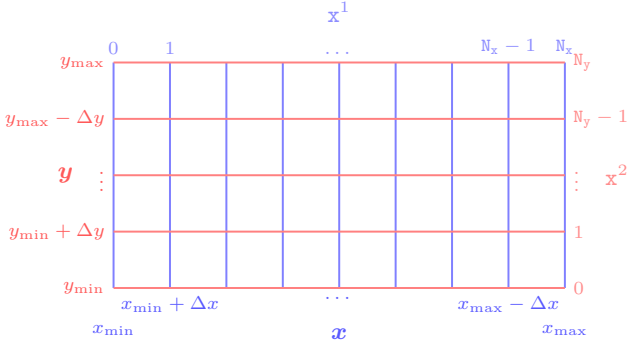

With this convention, $\Delta \mathtt{x^i}\equiv 1$, and the metric tensor has additional Jacobian terms corresponding to the effective ``stretching'' of the coordinates. Below, we present all the nonzero components of the metric in all three geometries discussed above.

\begin{itemize}
	\item Cartesian:
	      \begin{equation*}
		      \begin{aligned}
			      h_{11} & \equiv \left[(x_{\rm max}-x_{\rm min})/\mathtt{N_x}\right]^2,                       \\
			      h_{22} & \equiv \begin{cases}
				                      1~~~                                                     & \text{for 1D},    \\
				                      \left[(y_{\rm max}-y_{\rm min})/\mathtt{N_y}\right]^2~~~ & \text{for 2D/3D;}
			                      \end{cases} \\
			      h_{33} & \equiv \begin{cases}
				                      1~~~                                                     & \text{for 1D/2D}, \\
				                      \left[(z_{\rm max}-z_{\rm min})/\mathtt{N_z}\right]^2~~~ & \text{for 3D.}
			                      \end{cases}
		      \end{aligned}
	      \end{equation*}
	\item 2D axisymmetric spherical:
	      \begin{equation*}
		      \begin{aligned}
			      h_{11} & \equiv \left(\Delta r\right)^2,        \\
			      h_{22} & \equiv \left(r\Delta \theta \right)^2, \\
			      h_{33} & \equiv \left(r\sin{\theta}\right)^2,
		      \end{aligned}
	      \end{equation*}
	      where
	      \begin{equation*}
		      \begin{aligned}
			      \Delta r      & \equiv (r_{\rm max}-r_{\rm min})/\mathtt{N_r}, \\
			      \Delta \theta & \equiv \pi/\mathtt{N_\theta};                  \\
			      r             & \equiv r_{\rm min}+\mathtt{x^1}\Delta r,       \\
			      \theta        & \equiv \mathtt{x^2} \Delta\theta.
		      \end{aligned}
	      \end{equation*}
	\item 2D axisymmetric quasi-spherical:
	      \begin{equation*}
		      \begin{aligned}
			      h_{11} & \equiv \left(\Delta\mathcal{R} e^{\mathcal{R}}\right)^2,                                \\
			      h_{22} & \equiv \left(\Delta \mathcal{T} \partial_\mathcal{T}\theta[r_0+e^\mathcal{R}]\right)^2, \\
			      h_{33} & \equiv \left([r_0+e^\mathcal{R}]\sin{\theta}\right)^2,
		      \end{aligned}
	      \end{equation*}
	      where
	      \begin{equation*}
		      \begin{aligned}
			      \Delta \mathcal{R}         & \equiv  \left[\log{(r_{\rm max} - r_0)} - \log{(r_{\rm min} - r_0)}\right] / N_\mathcal{R}, \\
			      \Delta \mathcal{T}         & \equiv \pi/N_{\mathcal{T}};                                                                 \\
			      \mathcal{R}                & \equiv \log{(r_{\rm min} - r_0)} + \mathtt{x^1}\Delta\mathcal{R},                           \\
			      \mathcal{T}                & \equiv \mathtt{x^2}\Delta\mathcal{T};                                                       \\
			      \partial_\mathcal{T}\theta & \equiv  1 + 2 \vartheta +
			      12 \vartheta (\mathcal{T} / \pi)
			      (\mathcal{T} / \pi - 1),                                                                                                 \\
			      \theta                     & \equiv\mathcal{T}+2\vartheta \mathcal{T}(1-2\mathcal{T}/\pi)(1-\mathcal{T}/\pi).
		      \end{aligned}
	      \end{equation*}
\end{itemize}

The final ingredient we need are the coordinate conversion rules from the code-coordinate basis to the global orthonormal Cartesian basis: $\{\mathtt{x^1},\mathtt{x^2},\mathtt{x^3}\}\leftrightarrow\left\{x,y,z\right\}$.

\begin{itemize}
	\item Cartesian:
	      \begin{itemize}
		      \item forward:
		            \begin{equation*}
			            \{x,y,z\}=\{x,y,z\}_{\rm min}+\mathtt{x^{\{1,2,3\}}}\Delta\{x,y,z\}.
		            \end{equation*}
		      \item backward:
		            \begin{equation*}
			            \mathtt{x^{\{1,2,3\}}}=\left[\{x,y,z\}-\{x,y,z\}_{\rm min}\right]/\Delta\{x,y,z\}.
		            \end{equation*}
	      \end{itemize}
	\item 2D axisymmetric spherical:
	      \begin{itemize}
		      \item forward:
		            \begin{equation*}
			            \begin{aligned}
				            1. & ~r=r_{\rm min}+\mathtt{x^1}\Delta r,~\theta=\mathtt{x^2}\Delta \theta;
				            \\
				            2. & ~x = r \sin{\theta} \cos{\phi},
				            \\
				               & ~y=r\sin{\theta}\sin{\phi},
				            \\
				               & ~z=r\cos{\theta}.
			            \end{aligned}
		            \end{equation*}
		      \item backward:
		            \begin{equation*}
			            \begin{aligned}
				             & 1.~r=\sqrt{x^2+y^2+z^2},
				            \\
				             & ~~~~\theta=\pi/2-\text{atan2}\{z,\sqrt{x^2+y^2}\},
				            \\
				             & ~~~~\phi=\pi-\text{atan2}\{y, -x\};
				            \\
				             & 2.~\mathtt{x^1} = \left[r-r_{\rm min}\right]/\Delta r,
				            \\
				             & ~~~~\mathtt{x^2}=\theta/\Delta \theta.
			            \end{aligned}
		            \end{equation*}
	      \end{itemize}
	\item 2D axisymmetric quasi-spherical:
	      \begin{itemize}
		      \item forward:
		            \begin{equation*}
			            \begin{aligned}
				             & 1.~\mathcal{R}=\mathtt{x^1} \Delta\mathcal{R} + \mathcal{R}_{\rm min},
				            \\
				             & ~~~~\mathcal{T}=\mathtt{x^2}\Delta \mathcal{T};
				            \\
				             & 2.~r= r_0+e^{\mathcal{R}},
				            \\
				             & ~~~~\theta=\mathcal{T}+2\vartheta \mathcal{T}(1-2\mathcal{T}/\pi)(1-\mathcal{T}/\pi);
				            \\
				             & 3.~x = r \sin{\theta} \cos{\phi},
				            \\
				             & ~~~~y=r\sin{\theta}\sin{\phi},
				            \\
				             & ~~~~z=r\cos{\theta}.
			            \end{aligned}
		            \end{equation*}
		      \item backward:
		            \begin{equation*}
			            \begin{aligned}
				             & 1.~r=\sqrt{x^2+y^2+z^2},
				            \\
				             & ~~~~\theta=\pi/2-\text{atan2}\{z,\sqrt{x^2+y^2}\},
				            \\
				             & ~~~~\phi=\pi-\text{atan2}\{y, -x\};
				            \\
				             & 2.~\mathcal{R}= \log{(r-r_0)},
				            \\
				             & ~~~~\mathcal{T}=\{\text{see appendix~\ref{app:qspherical}}\};
				            \\
				             & 3.~\mathtt{x^1} = \left[\mathcal{R}-\mathcal{R}_{\rm min}\right]/\Delta \mathcal{R},
				            \\
				             & ~~~~\mathtt{x^2}=\mathcal{T}/\Delta \mathcal{T}.
			            \end{aligned}
		            \end{equation*}
	      \end{itemize}
\end{itemize}

In \entity, all of the metric tensor components, tetrad transformations, as well as the conversion rules to and from the global orthonormal Cartesian basis are methods provided within each \texttt{Metric} class. More technical details on this implementation are provided in later sections.

The duration of the timestep, $\Delta t$, in physical units is determined automatically by the code, using the user-provided value for the Courant-Friedrichs-Lewy (CFL) number. Namely, each individual \texttt{Metric} class computes the following quantity at initialization:

\begin{equation}
	\Delta x_{\rm min} = \min_{\texttt{i},\texttt{j},\texttt{k}}{\left(\sum_{i}^D\frac{1}{h_{ii}}\right)^{-1/2}},
\end{equation}

\noindent where the metric components are evaluated at the centers of each cell $(\texttt{\iPhalf},\texttt{\jPhalf},\texttt{\kPhalf})$. The timestep is then computed as $\Delta t = \texttt{CFL}\cdot\Delta x_{\rm min}/c$. For instance, in 2D Cartesian coordinates (with $\Delta x= \Delta y$), $\Delta x_{\rm min} = \Delta x / \sqrt{2}$, while in (quasi-)spherical coordinates, this value is determined by the size of the smallest cell.


\subsection{Digital filters for the currents}
\label{sec:filters}

As mentioned in the step-by-step algorithm outlined in Section~\ref{sec:time-discretization}, we apply consecutive digital filter passes to the deposited conformal currents, $\mathcal{J}^i$, before sourcing these to Amp\'ere's law. In the absence of boundary conditions, these filters have the following form:

\begin{itemize}
	\item 1D:
	      \begin{equation}
		      \mathcal{\tilde{J}}_{(\mathtt{i})}^i = \sum\limits_{-1\leq\{\mathtt{di}\}\leq1} f_{(\mathtt{di})}\mathcal{J}_{(\mathtt{i+di})}^i,
	      \end{equation}
	      where $f_{(\mathtt{di})} = (1/2)^{1+|\mathtt{di}|}$.
	\item 2D:
	      \begin{equation}
		      \label{eq:2d-filter}
		      \mathcal{\tilde{J}}_{(\mathtt{i},\mathtt{j})}^i = \sum\limits_{-1\leq \{\mathtt{di},\mathtt{dj}\}\leq1} f_{(\mathtt{di},\mathtt{dj})}\mathcal{J}_{(\mathtt{i+di}, \mathtt{j+dj})}^i,
	      \end{equation}
	      where $f_{(\mathtt{di}, \mathtt{dj})} = (1/2)^{2+|\mathtt{di}|+|\mathtt{dj}|}$.
	\item 3D:
	      \begin{equation}
		      \mathcal{\tilde{J}}_{(\mathtt{i},\mathtt{j})}^i = \sum\limits_{-1\leq \{\mathtt{di},\mathtt{dj},\mathtt{dk}\}\leq1} f_{(\mathtt{di},\mathtt{dj},\mathtt{dk})}\mathcal{J}_{(\mathtt{i+di}, \mathtt{j+dj},\mathtt{k+dk})}^i,
	      \end{equation}
	      where $f_{(\mathtt{di}, \mathtt{dj}, \mathtt{dk})} = (1/2)^{3+|\mathtt{di}|+|\mathtt{dj}|+|\mathtt{dk}|}$.
\end{itemize}

\noindent where, for brevity, we disregard the staggering of each individual component of the current when denoting the indices. The number of these filter passes is set by the input configuration at runtime. The filtered currents, $\mathcal{\tilde{J}}^i$, are then used as source terms in Amp\'ere's law. Note that near the polar axes in 2D (quasi-)spherical coordinates, as well as near the perfectly conducting boundaries, a special treatment is required for the currents. We address these cases respectively in Section~\ref{sec:axis} and Section~\ref{sec:pgen-shock}. Importantly, metric coefficients do not enter the filtering procedure, and the procedure can be used for an arbitrary metric and even in GR (as we show in the next paper).


\subsection{Special treatment of the polar axes}
\label{sec:axis}

Because of the coordinate singularity near the polar axes in 2D (quasi-)spherical coordinates, the denominator in \eqref{eq:discretized-ampere} for the evolution of $\AtNIJKtwoD{E}{n}{\iPhalf}{j}$ goes to zero. Because of this, for the components $E^1$, and $E^3$ in Amp\'ere's law, as well as the $B^2$ in Faraday's law, we treat the $\texttt{j}=0$ ($\theta=0$), and $\texttt{j} = \mathtt{N_2}$ ($\theta=\pi$) differently when (quasi-)spherical coordinates are employed. First of all, by symmetry, we set
\begin{equation}
	\begin{aligned}
		 & \AtNIJKtwoD{E^3}{n}{i}{\{0,N_2\}} = 0,       \\
		 & \AtNIJKtwoD{B^2}{n}{\iPhalf}{\{0,N_2\}} = 0,
	\end{aligned}
\end{equation}

\noindent as part of our boundary condition procedure, while skipping their update in the field-solver. Additionally, inside the Amp\'ere's loop, we use the integrated version of the equation for updating the $E^1$ component, similar to \cite{Belyaev_2015}:

\begin{equation}
	\begin{split}
		\Delta_{\texttt{n}}^{\texttt{n+1}}
		 & \left[\AtNIJKtwoD{E^1}{*}{\iPhalf}{j}\right]=                                                                                  \\
		 & \frac{\CovAtNIJKtwoD{B}{3}{n}{\iPhalf}{\jPhalf} - 2\pi \AtNIJKtwoD{\mathcal{J}^1}{\nPhalf}{\iPhalf}{j}}{A_{\texttt{\iPhalf}}},
	\end{split}
\end{equation}

\noindent Here, $A_\texttt{\iPhalf}\equiv A(\mathtt{x^1}=\texttt{\iPhalf})\equiv 2\pi \int_0^{1/2} \sqrt{h}~d\mathtt{x^2}$, and the additional factor of $1/2$ in $\mathcal{J}$ comes from the fact that the limits of integration in $\mathtt{x^2}$ are within half of the cell. Function $A$ can then be analytically defined for each metric:

\begin{itemize}
	\item 2D axisymmetric spherical:
	      \begin{equation*}
		      A(\mathtt{x^1})= r^2\Delta r(1 - \cos{\left\{\Delta\theta/2\right\}}),
	      \end{equation*}
	      where $r\equiv (r_{\rm min}+\mathtt{x^1}\Delta r)$.
	\item 2D axisymmetric quasi-spherical:
	      \begin{equation*}
		      A(\mathtt{x^1})=\left(r_0+e^{\mathcal{R}}\right)^2 e^{\mathcal{R}}\Delta \mathcal{R} (1 - \cos{\Delta\tilde{\theta}}),
	      \end{equation*}
	      where $\mathcal{R}\equiv \mathcal{R}_{\rm min}+\mathtt{x^1}\Delta\mathcal{R}$, and $\Delta\tilde{\theta}\equiv \Delta \mathcal{T}/2+\vartheta \Delta \mathcal{T}(1-\Delta \mathcal{T}/\pi)(1-\Delta \mathcal{T}/2\pi)$.
\end{itemize}

The filtering of conformal currents, described in Section~\ref{sec:filters}, also has to be modified near the polar axes. Considering the ``northern'' axis as an example (near $\theta = 0$), we impose the following conditions on the currents:
\begin{equation*}
	\begin{split}
		\mathcal{J}^1_{(\texttt{\iPhalf},\texttt{-1})} & = \mathcal{J}^1_{(\texttt{\iPhalf},\texttt{+1})}, \\
		\mathcal{J}^2_{(\texttt{i},\texttt{-1/2})}     & = -\mathcal{J}^2_{(\texttt{i},\texttt{+1/2})},    \\
		\mathcal{J}^3_{(\texttt{i},\texttt{0})}        & = 0,                                              \\
	\end{split}
\end{equation*}
\noindent which are used to compute the convolutions in \eqref{eq:2d-filter} (the corresponding coefficients, $f_{(\texttt{di},\texttt{dj})}$, remain the same).

\section{Structure of the code}
\label{sec:codestruct}

\entity~uses the \texttt{Kokkos} performance-portability library~\citep{Trott.etal_2022}, which allows the source code to be compiled on an arbitrary hardware architecture, internally translating all the constructs into \texttt{CUDA} (for NVIDIA GPUs), \texttt{HIP} (for AMD GPUs), and \texttt{SYCL} (for Intel GPUs) portability layers. In this section, we discuss how the internal logic of \entity~is adapted to take advantage of the hardware abstraction that \texttt{Kokkos} provides.

Aside from the standard set of problem generators (discussed later), \entity~also includes an extensive set of unit tests within the \texttt{CTest} framework, which are compiled separately, and check the validity of different aspects of the code on lower levels.


\subsection{Field and particle containers}

In the special relativistic module, \entity~stores 6 components of the electromagnetic fields, $E^i$, $B^i$, in a single $D+1$ dimensional array, where $D$ is the dimensionality of the problem; the last index corresponds to the particular component. E.g., for full 3D, the object containing field components will be defined as:

\begin{minted}{cpp}
Kokkos::View<real_t***[6]> EM;
// or defining an alias
ndfield_t<Dim::_3D> EM;
\end{minted}

\noindent where \texttt{real\_t} is compiled to either \texttt{float} (for single precision) or \texttt{double} (for double precision). We also used a template alias

\begin{minted}{cpp}
template<Dimension D>
ndfield_t<D> = ...;
// where
ndfield_t<Dim::_1D> 
// ... is same as Kokkos::View<real_t*[6]>
ndfield_t<Dim::_2D> 
// ... is same as Kokkos::View<real_t**[6]>
ndfield_t<Dim::_3D> 
// ... is same as Kokkos::View<real_t***[6]>
\end{minted}

\noindent The type \texttt{Dimension} here corresponds to a simple \texttt{unsigned short}, i.e., \texttt{Dim::\_1D == (unsigned short)1}, etc.

The size of the \texttt{EM} array in spatial dimensions corresponds to the number of cells plus the number of \textbf{ghost cells}, \texttt{N\_GHOST}, in both positive and negative direction. This number is a compile-time parameter, and can be adjusted to meet specific needs (the default value used is $\texttt{N\_GHOST=2}$). Since the indexing of the array begins at \texttt{(0,0,0)}, the first active cell is located in memory at the index \texttt{(N\_GHOST,N\_GHOST,N\_GHOST)}. With this notation, to access, for instance, $B^2_{(\texttt{i+1/2,j,k+1/2})}$, we simply take

\begin{minted}{cpp}
EM(N_GHOST + i, N_GHOST + j, N_GHOST + k, 
   em::bx2)
\end{minted}

\noindent where we also define simple integer aliases to quickly access the specific components of the field

\begin{minted}{cpp}
enum em { ex1 = 0, ex2 = 1, ex3 = 2, 
          bx1 = 3, bx2 = 4, bx3 = 5 };
\end{minted}

Particle species in \entity~are stored as a standard C++ vector of structures containing all the information about each individual particle:

\begin{minted}{cpp}
std::vector<Particles<D>> species;
\end{minted}

\noindent Each of the \texttt{Particles<D>} structures store individual particle quantities, such as their positions, velocities, etc., as 1D \texttt{Kokkos::View}-s, as well as the mass and charge of each individual species, as well as the allocation size and the number of active particles (discussed later):

\begin{minted}{cpp}
template<typename T>
using array_t = Kokkos::View<T*>;

template<typename T>
using array2d_t = Kokkos::View<T**>;

using npart_t = std::uint64_t;

template<Dimension D>
Particles {
    float mass, charge;
    npart_t npart, maxnpart;
    
    array_t<int> i1, i2, i3,
                 i1_prev, i2_prev, i3_prev;
    array_t<prtldx_t> dx1, dx2, dx3,
                      dx1_prev, dx2_prev, 
                      dx3_prev;
    
    array_t<real_t> phi;

    array_t<real_t> ux1, ux2, ux3;
    array_t<real_t> weight;
    array_t<short> tag;
    array2d_t<real_t> pld_r;
    array2d_t<npart_t> pld_i;
};
\end{minted}

\noindent here we also used a template alias \texttt{array\_t} for brevity, as well as alias types \texttt{prtldx\_t}, which is typically compiled as \texttt{float} (but can potentially be used in double precision), and \texttt{npart\_t}, which is normally aliased to \texttt{unsigned} or \texttt{unsigned long}.

Notice that the coordinate of each particle in each direction is stored as a combination of two quantities: \texttt{i1} --- the index of the cell the particle is located in, and \texttt{dx1} ($\in[0,1)$) --- the displacement of the particle within that cell. This choice has two separate advantages. Firstly, it avoids the loss of accuracy when the range of positions varies by several orders of magnitude; secondly, it allows to avoid computing the full position of the particle in routines where only the displacement within a cell is needed (e.g., field interpolation or current deposition). Also, notice that we store the previous position of the particle, as discussed above, which is used when depositing the electric currents. In 1D/2D simulations only the necessary arrays are allocated (based on the value of the template argument \texttt{D}), i.e., in Cartesian 2D, we do not allocate \texttt{i3}, \texttt{i3\_prev}, \texttt{dx3}, \texttt{dx3\_prev}. Also, as discussed above, in 2D axisymmetric spherical coordinates we keep track of particles' $\phi$ coordinates with the dedicated array --- \texttt{phi}. Particles of each species can also be prescribed an arbitrary number of custom payloads (both real-valued as well as integers), both of which are defined as 2D arrays (the first index being the particle's index, and the second one being an identifier for the given payload). These payloads can be used for a variety of custom purposes, such as the storage of trajectory-integrated quantities, custom counters, etc.

Each of the necessary particle arrays is pre-allocated at the beginning of the simulation using the user-provided \texttt{maxnpart} number, and the number of actively used particles is kept track of via the \texttt{npart} variable (all the particle arrays go from \texttt{0} to \texttt{npart}). The \texttt{tag} variables store a \texttt{short}-valued quantity for each particle, indicating whether the particle is (a) alive, (b) scheduled for deletion, or (c) scheduled to be sent through the boundaries of the domain. Thus, any loop over all the particles has the following form:

\begin{minted}{cpp}
for (auto& particles : species) {
  Kokkos::parallel_for("ParticleLoop", 
    particles.npart,
    KOKKOS_LAMBDA(const npart_t p) {
      if (particles.tag(p) != Tag::Alive) {
        return;
      }
      // action with the p-th particle
    }
  );
}
\end{minted}

\noindent where \texttt{Tag::Alive} is aliased to \texttt{(short)1}. Note that the external loop over all the species is performed on the CPU (serially), while the loop over each individual particle in the given species is vectorized (potentially, on GPUs).



\subsection{Kernels}

Kernels in \entity~are special functor classes (i.e., classes with an overloaded \texttt{operator()} method) which implement all of the algorithms discussed in Section~\ref{sec:curvspace}. There are three kernel archetypes which roughly correspond to: (a) field-solvers, (b) particle-pushers, and (c) current depositions. The first two are trivially vectorizable, since the operations on all the distinct elements are independent of each other. The current deposition algorithm relies on accumulating the electric currents from each particle onto the grid, and thus potentially involves race conditions when two particles from close-by attempt to write into the same cell independently. Below, we discuss briefly how each of these kernels is implemented.

First of all, for the field-solver kernels (both the Faraday and Amp\'ere push for the fields), we implement both Cartesian as well as curvilinear routines separately, since in Cartesian, equations~\eqref{eq:discretized-faraday} and~\eqref{eq:discretized-ampere} can be largely simplified, especially if one further assumes that $\Delta x=\Delta y=\Delta z$. The Cartesian version of the fieldsolver kernels has the following general form:

\begin{minted}{cpp}
namespace kernels::mink {

template<Dimension D>
class Ampere_kernel {
  ndfield_t<D> EB;

  const real_t c1, c2;
  
  KOKKOS_INLINE 
  void operator()(index_t i1) const {
    // 1D implementation
    //   c1 = dt / dx
    // EB(i1, em::) += c1 * curl(EB(i1, em::));
  }
  KOKKOS_INLINE 
  void operator()(index_t i1, 
                  index_t i2) const {
    // 2D implementation
    //   c1 = dt / dx^2; c2 = dt
    // EB(i1, em::) += {c1,c2} * 
    //                  curl(EB(i1, em::));
  }
};

}
\end{minted}

\noindent Notice that here the constant coefficients \texttt{c1} and \texttt{c2} are precomputed and passed to the kernel. Non-cartesian kernels, on the other hand, use the $h_{ij}$ components from the metric object passed to the kernel:

\begin{minted}{cpp}
namespace kernels::sr {

template<class M>
class Ampere_kernel {
  ndfield_t<M::Dim> EB;
  const M metric;

  const real_t dt;
  
  KOKKOS_INLINE 
  void operator()(index_t i1, 
                  index_t i2) const {
    // 2D implementation
    const auto one_ovr_sqrt_detH_pH0 = 1 / 
      metric.sqrt_det_h({i1 + 1/2, i2});
    const auto h3_pHpH = 
      metric.template h_<3, 3>({i1 + 1/2, 
                                i2 + 1/2});
    const auto h3_pHmH = 
      metric.template h_<3, 3>({i1 + 1/2, 
                                i2 - 1/2});
    EB(i1, i2, em::ex1) += dt * 
      one_ovr_sqrt_detH_pH0 *
      (h3_pHpH * EB(i1, i2, em::bx3) -
       h3_pHmH * EB(i1, i2 - 1, em::bx3));
    // ...
  }
};

}
\end{minted}

\noindent The metric typically takes in the coordinates (in code units) or vectors of types, respectively, \texttt{coord\_t<D>} and \texttt{vec\_t<D>}, both of which are aliased to \texttt{real\_t[D]}.

Particle-pusher kernel is conceptually also simple, as discussed in the previous section. It accepts individually all the relevant 1D arrays for the given species, i.e., \texttt{i1}, \texttt{dx1}, \texttt{ux1}, etc, as well as the electromagnetic fields. Before updating the position, the kernel also records the previous position of the particle: \texttt{i1\_prev(p)=i1(p)} and \texttt{dx1\_prev(p)=dx1(p)}, etc.

The current deposition kernel takes advantage of \texttt{Kokkos}'s built-in \texttt{ScatterView}'s to avoid race conditions when particles attempt to deposit current into the same grid point. \texttt{ScatterView} data structure internally employs hardware-level atomics in most situations, while in certain cases (low thread count on CPUs) it uses data replication. The entire current deposit routine has the following form:\footnote{The current deposition algorithm itself (for the first order shape function) uses a modified version of the zig-zag algorithm by \cite{Umeda.etal_2003}, computing the intermediate point of the trajectory in such a way that conversion from \texttt{\{i,di\}} to real-valued coordinate, \texttt{(real\_t)(i+di)}, is never necessary.}

\begin{minted}{cpp}
namespace KE = Kokkos::Experimental;
auto scatter_J = KE::create_scatter_view(J);
for (auto& particles : species) {
  Kokkos::parallel_for("CurrentsDeposit",
    particles.npart,
    KOKKOS_LAMBDA(index_t p) {
      // precompute shape functions
      auto J_acc = scatter_J.access();
      J_acc(...) = ...;
    }
  );
}
KE::contribute(J, scatter_J);
\end{minted}

\noindent Where the currents are stored in the same way as the electric and magnetic field, except the number of components is $3$ instead of $6$.


\subsection{Communications}

To allow efficient parallelization across multiple devices, the entire domain of the simulation is subdivided into multiple subdomains. These are typically distributed on a logically cartesian grid in such a way that the number of cells on each is roughly the same. The total number of subdomains, $s$, can thus be expressed as a product of numbers in each direction -- for 3D this would be $s_1\times s_2 \times s_3$. These three numbers are either specified by the user in the input file during the start of the simulation, or can also be inferred by the code internally, provided the total number $s$. The user can also opt to specify only some of the $s_i$ numbers, in which case the others, which should be set to $s_j = -1$, will be picked automatically. For example, the user might pick $s_i=\{2,-1,-1\}$, which means that $s_1=2$ always, while the other two will be determined by the code automatically, given that $s_2\times s_3=s/2$, to best balance the number of cells on each subdomain.

Each subdomain, thus, only allocates fields to represent a subset of the grid. Note, that the metric on each of the subdomains is different; this should come as no surprise, since the coordinate system on each subdomain is independent, and the grid in the $i$-th direction goes from $0$ to $\mathtt{N}_i$, where $\mathtt{N}_i$ is the number of cells on that specific subdomain. Thus, in general, when transferring particles from one subdomain to the other, it is necessary to convert their coordinates. Conveniently, this is rather trivial to do, as only the integer part of the coordinate is transformed, while the displacement within the cell remains the same. Four-vectors, on the other hand, should not be transformed, since the metric components themselves are continuous throughout the subdomains.

The parallelization of the tasks is done with the \texttt{MPI} library, where each subdomain is then assigned a specific \texttt{mpi\_rank}. Depending on the boundary conditions, each subdomain contains information about its neighboring subdomains (in particular, the \texttt{mpi\_rank}-s of each of the neighbors): in 1D, there are only $2$ neighbors; in 2D, the number of neighbors is $8$, while in 3D -- it's $26$. In case the subdomain is at the edge of the box with non-periodic boundaries, the corresponding direction will have a \texttt{nullptr} instead of the neighbor.

\subsubsection{Fields \& currents}

Since the field solvers as well as the field interpolations on particle positions have a finite stencil and thus potentially require components outside of the current subdomain, we store an additional layer of cells -- the \emph{ghost zone} -- around the edges of each of the subdomains.\footnote{This means that the shapes of the field components on each of the subdomains in each direction are \texttt{(n\_i + 2 * N\_GHOSTS)}.} These extra ghost layers are ``filled'' by communicating each individual subdomain with its neighbors each time one of the field or current components is updated (i.e., after the Faraday/Amp\'ere steps, or the current deposit).

The communication loop itself is performed in the following way (the same logic is also used further when communicating particles):

\begin{minted}{cpp}
for (auto dir : all_directions) {
  auto nbr_to_send = domain.neighbor_in(dir);
  auto nbr_to_recv = domain.neighbor_in(-dir);
  if (nbr_to_send == nullptr and 
      nbr_to_recv == nullptr) {
    continue;
  } else if (nbr_to_send == nullptr) {
    MPI_Recv(..., nbr_to_send.rank, ...);
  } else if (nbr_to_recv == nullptr) {
    MPI_Send(..., nbr_to_recv.rank, ...);
  } else {
    MPI_SendRecv(...,
      nbr_to_send.rank,
      nbr_to_recv.rank, 
      ...);
  }
}
\end{minted}

\noindent Since the calls to \texttt{MPI\_\{Send,Recv\}} are blocking (i.e., the code will not proceed further unless the data has been both sent and received), the method of sending in one direction, while receiving from the opposite one, ensures that each time a rank is sending data, there is a corresponding rank expecting to receive either via \texttt{MPI\_Recv} or \texttt{MPI\_SendRecv}.

The electric and magnetic fields, as well as the electric currents, are communicated via first copying the data into temporary small buffers (of the size of the corresponding ghost zone). Since some systems do not have support for GPU-aware \texttt{MPI}, \entity~provides a compile-time flag to disable the direct communication between the devices, instead opting to communicate indirectly via copying to the CPU. For example, the implementation of the \texttt{MPI\_Recv} function looks like the following:

\begin{minted}{cpp}
#if !defined(DEVICE_ENABLED) || \
     defined(GPU_AWARE_MPI)
  MPI_Recv(recv_arr.data(), ...);
#else
  auto recv_arr_h = 
    Kokkos::create_mirror_view(recv_arr);
  MPI_Recv(recv_arr_h.data(), ...);
  Kokkos::deep_copy(recv_arr, recv_arr_h);
#endif
\end{minted}

\noindent Here, in the second case, we first create the so-called \emph{host-mirror} of the original array (which is basically a copy of the original array on the CPU), and receive into it, before copying it back to the device memory. The \texttt{.data()} simply returns the raw pointer to the data stored in the corresponding \texttt{Kokkos::View}.

There are, in general, two types of communications for the electric currents. In the first case, just like for the electric and magnetic fields, we fill the ghost zones on the current subdomain with the data from the corresponding neighboring subdomains. In the second type, which we refer to as the \emph{currents synchronization}, and which is performed once after the current deposition step, the currents deposited by particles into the ghost zone of the current subdomain are sent and added into the active zone of the neighboring domains. To avoid double-counting, this data is first accumulated into an empty buffer, which is added to the main currents array only after synchronization in all directions has been performed.

\subsubsection{Particles}

As particles cross the boundaries between the subdomains, they too need to be redistributed from the current domain to its neighbors. This operation is performed in multiple steps. First, within the particle pusher kernel, the particle's \texttt{tag} variable is changed to indicate in which direction the particle is crossing the current subdomain. Each direction has a corresponding tag (which is simply a \texttt{short}-valued number \texttt{>1}), which, for simplicity, we further denote as \texttt{Tag::SendIn(dir)}, to indicate the particle is being sent to direction \texttt{dir}. In addition, \texttt{Tag::Dead=0} indicates that the particle should simply be deleted, while \texttt{Tag::Alive=1} indicates that no additional communication is required for the particle. After particles are tagged in the pusher, and after the current deposition substep is executed, the communication routine is called. Below, we outline all the steps involved in sending/receiving particles to/from the neighboring subdomains.

\begin{enumerate}
	\item Construct an array, \texttt{npptag}, containing the number of particles having a specific tag; i.e., \texttt{npptag(Tag::SendIn(dir))} indicates the total number of particles being sent in the specific direction \texttt{dir}. Also compile an array of offsets (since we will buffer all the to-be-sent or -deleted particles in a single array), where
	      \begin{equation*}
		      \texttt{offset(tag)} = \sum\limits_{\texttt{t}<\texttt{tag}}^{\texttt{t}\ne \texttt{Tag::Alive}} \texttt{npptag(t)}.
	      \end{equation*}
	\item Record the indices for all the particles that have \texttt{tag != Tag::Alive}, into a 1D buffer array, \texttt{dead\_indices}, of the size
	      \begin{equation*}
		      \sum\limits_{\texttt{t}\ne \texttt{Tag::Alive}} \texttt{npptag(t)},
	      \end{equation*}
	      where the \texttt{offset} array is used to clearly separate particles into corresponding groups.
	\item Update the \texttt{i1}, \texttt{i2}, \texttt{i3}, \texttt{i1\_prev}, \texttt{i2\_prev}, \texttt{i3\_prev} positions of the to-be-sent particles to account for their positions in the new subdomain.
	\item For the to-be-sent particles, all the quantities with specific types, such as \texttt{int}, \texttt{real\_t}, \texttt{prtldx\_t}, are then stored in the dedicated buffers for each type, with \texttt{i1}, \texttt{i2}, ... being stored in \texttt{buff\_int}; \texttt{dx1}, ... -- in \texttt{buff\_prtldx}; \texttt{ux1}, ... -- in \texttt{buff\_real}. Since we have already computed the number of particles to-be-sent in each direction, we can easily recover the sub-arrays that need to be sent in each individual direction. All the tags for these particles on the current subdomain are then set to \texttt{Tag::Dead}.
	\item At the same time, we also pre-allocate buffers for each type to store the received particles from all of the neighbors.
	\item When the send/receive operations for all directions are performed, we may extract the received quantities stored in the dedicated buffers, overwriting the dead/sent particles using the previously compiled \texttt{dead\_indices} array. If the number of received particles exceeds the number of particles that have either ``died'' or were sent to the neighboring subdomains, we simply place them at the end of the array, simultaneously increasing the \texttt{npart} parameter.
\end{enumerate}

Since the positions of newly ``dead'' particles are being reused to place the newly received ones from the neighboring domains, this algorithm has the advantage of minimizing the number of ``holes,'' i.e., the number of particles with \texttt{tag(p) == Tag::Dead}, and whose index is \texttt{p < npart}. Nonetheless, to ensure the ``holes'' are being cleared out consistently throughout the simulation, once every few hundred timesteps (a parameter controlled by the user) we call an additional routine which sorts all the particle quantities based on their \texttt{tag}, after which \texttt{npart} always indicates the number of ``alive'' particles, while the ``dead'' ones are shuffled to the end of the allocated arrays (positions \texttt{p = npart} to \texttt{p = maxnpart}).


\subsection{Problem generators}
\label{sec:problem_generators}

The main interface through which the user interacts with \entity~ is via the so-called problem generators. These are essentially a collection of functions and routines bundled into a template class called \texttt{PGen}. In most use-cases, the user can control everything about the simulation through the problem generator, and no additional edits in the source code are required. The typical problem generator class looks like the following:

\begin{minted}{cpp}
template <SimEngine S, class M>
struct PGen {
  // init of fields through a custom class:
  MyInitFields<D> init_flds;

  // particle initialization"
  inline void InitPrtls(Domain<S, M>&) {
    // call particle injector
  }
  
  // function called after each timestep:
  inline void CustomPostStep(
    timestep_t,
    simtime_t,
    Domain<S, M>&
  ) {
    // e.g., inject particles,
    //   apply custom boundaries, etc.
  }

  /*
   * other parameters/routines:
  **/
  // external current source
  MyExtraCurrents<D> ext_current;

  // custom matching field boundaries:
  auto MatchFields(real_t) const 
    -> MyMatchFields<D> {
    // return custom class 
    //   to be used in matching boundaries
  }
  // etc...
};
\end{minted}

\noindent Note that none of the methods or internal parameters mentioned above are strictly necessary, as the code checks for their existence at compile time before calling or addressing. However, the naming conventions must be respected for the compiler to find the proper variables or methods; e.g., the structure containing the initial fields has to be called \texttt{init\_flds} (the name of the actual class template -- in this case, \texttt{MyInitFields} -- is irrelevant, and can be anything), while the method that initializes the particles has to be called \texttt{InitPrtls}, and the method called at the end of each timestep has to be called \texttt{CustomPostStep}. The capabilities of the problem generator class are quite big, and we will not go over all of its aspects (they are outlined in more detail in our \href{https://entity-toolkit.github.io/wiki/content/2-howto/1-problem_generators/}{documentation}\footnote{\url{https://entity-toolkit.github.io/wiki/content/2-howto/1-problem_generators/}}).

\subsubsection{User-defined fields \& currents}

The problem generator often needs to provide user-defined fields or electric currents, like in the example shown above, where the initial fields are contained within the variable called \texttt{init\_flds} (mandatory name) having a type of \texttt{MyInitFields<D>} (arbitrary name). That class has to have methods for defining one or more of the electric or magnetic field components:

\begin{minted}{cpp}
template <Dimension D>
struct MyInitFields {
  const real_t Delta, Xc;
  MyInitFields(real_t delta, real_t xc) 
  : Delta { delta }, Xc { xc } {}

  KOKKOS_INLINE
  auto bx2(const coord_t<D>& x) const 
    -> real_t {
    return math::tanh((x[0] - Xc) / Delta);
  }

  KOKKOS_INLINE
  auto ex3(const coord_t<D>&) const 
    -> real_t {
    return 0.25;
  }
};
\end{minted}

\noindent In this example, the class takes in two external parameters, $\Delta=\texttt{Delta}$, and $x_c=\texttt{Xc}$, and provides rules for initializing $B^{\hat{1}}=\tanh{\{(x-x_c)/\Delta\}}$, and $E^{\hat{3}}=0.25$; all components of the fields are defined in the tetrad (orthonormal) basis, and are normalized to fiducial values (see Section~\ref{sec:units}). The coordinates are in the global ``physical'' basis; their values range within the physical extent of the entire box, e.g., $x_{\rm min}\to x_{\rm max}$, for Cartesian or $r_{\rm min}\to r_{\rm max}$ for (quasi-)spherical.

Notice that in this example we have not defined all the components, which will, as a result, be ignored. The same logic applies when the problem generator specifies target fields for the matching boundaries conditions, or external currents (both discussed below; in the case of external currents, the methods are called \texttt{jx1}, \texttt{jx2}, and \texttt{jx3}).

\subsubsection{Particle injection}

\entity~ provides a variety of pre-fabricated routines for particle injection for the most common use-cases. Two of the most important ones are the \emph{uniform} and the \emph{non-uniform} injector. The main difference between the two is the way the underlying kernel parallelizes its execution. In the former case, the number of injected particles is known in advance, and thus the procedure is vectorized across the injected particles. In the latter case, the loop is vectorized across the cells of the grid, where at each point a specific function is computed, estimating the number of locally injected particles. The indices of particles are then updated atomically. In both cases, the code injects two particle species of opposing charges simultaneously (the charges are injected on top of each other to avoid initializing non-zero charge density), while the velocity distribution is determined by a specific class provided to the injector.

A typical call to a uniform injector looks like the following:

\begin{minted}{cpp}
// initialize the velocity distribution
//   uses the built-in Maxwellian
const auto vel_dist = 
  Maxwellian(
    metric, random_pool, temperature
  );
// call the uniform injector
InjectUniform(
  params, domain, 
  { 1, 2 }, // indices of species to inject
  { vel_dist, vel_dist } // velocity distribution 
                         // for each species
  tot_density
);
\end{minted}

One may also easily define a custom velocity distribution functor, where the \texttt{operator()} has to accept the coordinate, \texttt{const coord\_t<D>\&}, the velocity to-be-set, \texttt{vec\_t<Dim::\_3D>\&}, and the species index, \texttt{unsigned short}.

Likewise, the non-uniform injector has an additional parameter that defines the spatial distribution of the injected plasma:

\begin{minted}{cpp}
template <SimEngine S, class M>
struct DensityProfile {
  KOKKOS_INLINE
  auto operator()(const coord_t<M::Dim>& x) 
    const -> real_t {
    const auto r_sqr = 
      x[0] * x[0] + x[1] * x[1];
    return math::exp(-r_sqr / 10.0);
  }
};

const auto spat_dist = DensityProfile();
InjectNonUniform(
  params, domain, 
  { 1, 2 }, 
  { vel_dist, vel_dist },
  spat_dist, // spatial distribution functor
  max_tot_density
);
\end{minted}

\noindent where the value returned by the \texttt{DensityProfile} functor (typically between $0$ and $1$) will then be multiplied by \texttt{max\_tot\_density} to determine the locally injected density.


\subsection{Data output \& post-processing}

\entity~supports multiple types of data output for further post-processing of the simulation results: grid-based fields, individual particles, and energy distribution functions. We use the highly-scalable \texttt{ADIOS2} library~\citep{ADIOS2} which supports multiple data output formats, including \texttt{BP5}, and \texttt{HDF5} (with the former typically performing much faster on large numbers of GPUs). For each output category (fields, particles, etc), the output cadence can be configured separately in the input file, and the files for each timestep are stored separately into corresponding directories.

Of the grid-based quantities, we output the electric and magnetic field, and the electric current components interpolated to the center of each cell; each value is also converted into a tetrad (orthonormal) basis and normalized to $B_0$ and $4\pi q_0 n_0$, respectively. One may also output the divergence of the electric field, $\texttt{divE}$, computed in the nodes. Additionally, we also support outputting particle moments on the grid, in particular the number density, $\texttt{N\_s}$ (normalized to $n_0$), the mass and charge density, $\texttt{Rho\_s}$ and $\texttt{Charge\_s}$  (normalized to $m_0n_0$ and $4\pi q_0 n_0$ respectively),  the mean three-velocity, $\texttt{V\_s}$ (defined as $\bm{V}\equiv \int d^3\bm{u}f(\bm{u})\bm{u} / u^0$),  the number of particles per each cell $\texttt{Nppc\_s}$, as well as all the components of the stress-energy tensor $\texttt{Tij\_s}$, defined as $T^{\mu\nu}\equiv\int d^3\bm{u}f(\bm{u})u^\mu u^\nu/u^0$. Here, the ``$\texttt{\_s}$'' notation indicates that the moment is computed using the species $s$; e.g., to compute the number density of the first and the second species combined, one could use $\texttt{N\_1\_2}$, or to compute the $T^{0i}$ for just the third species: $\texttt{T0i\_3}$. We also support outputting custom grid-based quantities, which can easily be defined within the used problem generator. For the particles, we output their positions (in global physical coordinates), four-velocities, payloads, and unique id-s (for which one has to set \texttt{tracking = true} for the given species). Both the grid-based quantities as well as particles can be downsampled for the output to reduce the storage usage. \entity~also supports an output of specific box-averaged scalars and vectors, such as the average $E^2$, $B^2$, $\bm{J}\cdot\bm{E}$ components of $\bm{E}\times\bm{B}$, etc, which can be used to quickly probe the energy/momentum conservation.

\entity~also comes with a Python package for post-processing the data -- \texttt{nt2py}\footnote{Explicit link: \href{https://pypi.org/project/nt2py}{https://pypi.org/project/nt2py}}. This package, which works both for the \texttt{BP5} and the \texttt{HDF5} formats, uses \texttt{dask} and \texttt{xarray} Python packages to provide a lazy-loading interface and quick, intuitive access to specific chunks of the data. The typical usage of the package looks as follows:

\begin{minted}{python}
import nt2
data = nt2.Data("MySimulation")

data.fields.N\
    .sel(t=12.3, method="nearest")\
    .sel(x=slice(5, 8))\
    .mean("y")\
    .plot()
\end{minted}

The example code above lazily loads the simulation data, accesses the number density, \texttt{N}, selects the time $t\approx12.3$ (\texttt{method="nearest"} means it will pick the closest time to the indicated one), selects an interval $5\leq x<8$, averages the quantity in $y$, and the plots the result. The result is either a 1D or a 2D plot, depending on the original dimensionality of the data (since we average the quantity along one of the dimensions). The advantage of lazy loading is that the code will automatically load into memory only the data chunks necessary for performing the described operation, which avoids out-of-memory issues with specifically large datasets.

\section{Performance}
\label{sec:performance}


Performance portability of \entity~relies heavily on the \texttt{Kokkos} library, which allows to compile and run the code on various GPU micro-architectures across different vendors. We have successfully tested the code on some of the major national supercomputers in both the US and Europe (instructions on setting up the dependencies are readily available on \href{https://entity-toolkit.github.io/wiki/content/useful/cluster-setups/}{our documentation website}\footnote{Explicit link: \url{https://entity-toolkit.github.io/wiki/content/useful/cluster-setups/}.}).

\begin{figure}
	\centering
	\includegraphics[width=\columnwidth]{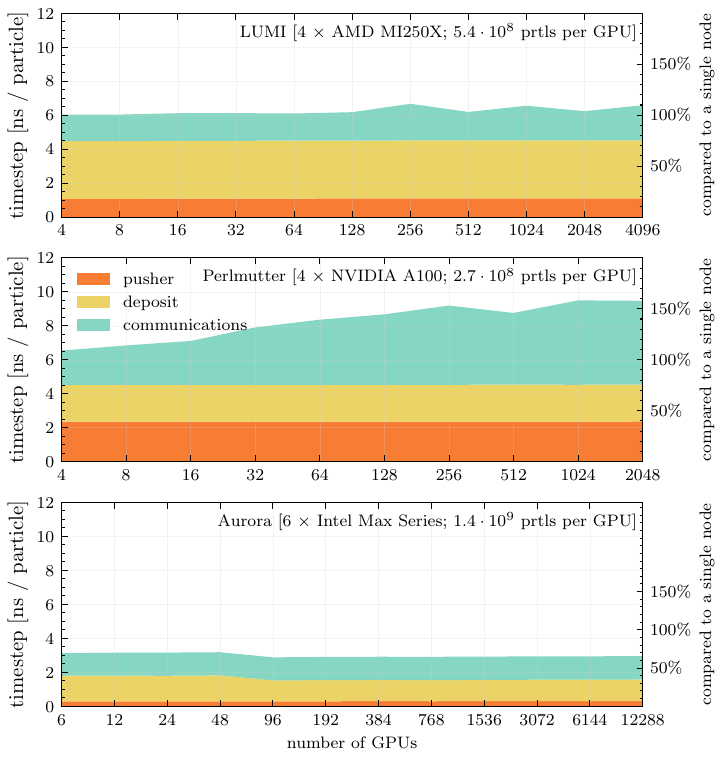}
	\caption{Stacked plot showing the weak scaling performed by simulating the 3D Weibel instability in a periodic box. The size of each meshblock, as well as the total number of particles, is kept constant per each GPU (or each device unit), i.e., the total size of the box and the number of particles are increased proportionally to the number of GPUs. The name of the supercomputer, the GPU node architecture, as well as the total number of particles per GPU are indicated at the top right of each panel.}
	\label{fig:weak_scaling}
\end{figure}

Below, we present the results from weak scaling test runs on three major representative platforms: AMD MI250X (LUMI cluster in Finland), NVIDIA A100 (Perlmutter, NERSC, US), and Intel Max Series (Aurora, ANL, US). For each respective architecture, \texttt{Kokkos} automatically uses corresponding backends: HIP on AMD, CUDA on NVIDIA, and SYCL on Intel GPUs. For parallelization, we use the OpenMPI library on Perlmutter, and MPICH -- on LUMI \& Aurora; additionally, on LUMI and Perlmutter, because of hardware limitations, we disable the GPU-aware-MPI option, forcing the code to communicate the data between meshblocks via the CPU. For the test, we use a 3D single-precision simulation of the Weibel instability in a periodic box (setup is discussed in Section~\ref{sec:pgen_streaming}), where the total number of cells and the number of particles is increased proportional to the number of GPUs used (on LUMI, and Perlmutter, we use $256^3$ cells with $16$ particles per cell, while on Aurora, because of its larger VRAM, we take $256^2\times 341$ cells with $64$ particles per cell). On AMD MI250X, we used two MPI ranks per GPU, one per each graphics compute die (GCD), but for simplicity of cross-comparison, we normalize the performance per GPU, instead of per rank. Similarly, for Intel Max Series, we use two MPI ranks per GPU, one per subdomain. The performance breakdown for all three architectures vs the number of GPUs used is demonstrated in Figure~\ref{fig:weak_scaling}. In all cases, the fieldsolver, currents filtering, etc. take less than a few percent of the entire timestep duration combined, and we thus only report three of the most expensive substeps -- communications, particle pusher, and current deposition -- quantifying the performance in units of nanoseconds per particle per each GPU per timestep. In all three architectures, the particle pusher takes about $2$ nanoseconds per particle (in 1D and 2D cases not reported here, this number can be as low as about $0.5...1$ nanosecond). Current deposition relies on atomic operations performed on a global per-meshblock array of electric currents; despite that, on both NVIDIA and Intel architectures, this operation is either comparable or even faster than the particle pusher, while on AMD the deposit substep is about two times slower than the pusher. The surprisingly strong performance boost on Aurora's Intel chips -- especially for the particle pusher (which performs random access to the field components on the grid, and yet on Intel takes, on average, about $0.3$ ns per particle each timestep) -- is likely due to the large cache size the Intel's PVC chips have ($\sim 400$ MB for the L2 level cache).

The communications substep, which involves all the data transfer between the meshblocks, including all the field components, the currents, as well as the particles, takes on average about two-to-three times less than the deposit and the pusher combined (i.e., $25\%...30\%$ of the duration of the entire timestep). For the Perlmutter, the communication performance, which has been forced to ignore GPU-awareness, drops twice from $4$ to $2048$ GPUs, but at most takes about half of the duration of the entire timestep.

\begin{figure}
	\centering
	\includegraphics[width=\columnwidth]{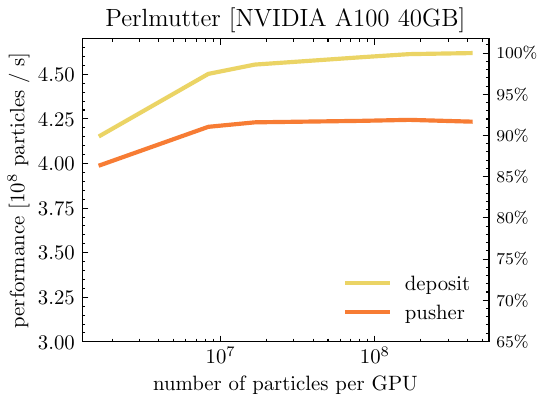}
	\caption{Duration of the current deposition and the particle pusher substeps as a function of the total number of particles per GPU on the Perlmutter NVIDIA A100 GPU with 40 GB of VRAM (strong scaling).}
	\label{fig:strong_scaling}
\end{figure}

While this test evaluates the best-case scenario in which a large fraction of the GPU memory is being utilized, oftentimes the loads for each GPU may be heavily misbalanced, and some of the meshblocks may contain significantly fewer particles than others. To test this effect, in Figure~\ref{fig:strong_scaling} we show the result of a strong scaling, where with the same setup on a single GPU node, we fix the size of the domain and vary the total number of particles (the maximum value is capped by the 40 GB VRAM of the NVIDIA A100 GPU on Perlmutter). Evidently, when dropping the number of particles to about $1\%$ of the total memory capacity, the performance per particle is degraded by only about $10\%$.

\begin{figure}
	\centering
	\includegraphics[width=0.8\columnwidth]{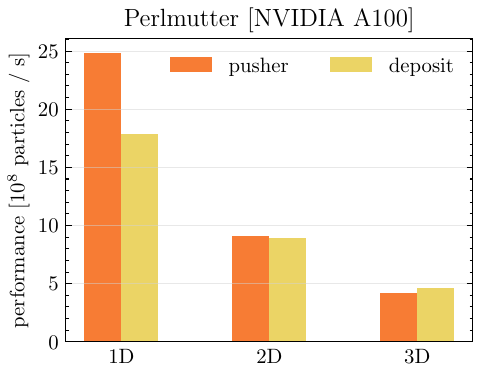}
	\caption{Same experiment as in Figure~\ref{fig:strong_scaling}, except the number of particles and cells is kept constant, while we change the dimensionality of the problem.}
	\label{fig:performance_dim}
\end{figure}

It is also worth noting that the dimensionality of the problem has a significant effect on the performance, as the kernels both access less memory for lower-dimensional problems, thus enabling more efficient data caching on the threads, and also the number of operations is slightly reduced. This is demonstrated in Figure~\ref{fig:performance_dim}, where we do the same experiment as before, except now the number of particles is kept fixed in 1D, 2D, and 3D, while the box size is adjusted to ensure the same number of cells in all cases. The particle pusher kernel, which is primarily memory bandwidth bound in 3D, speeds up by a factor of $\times 2$ in 2D and almost by a factor of $\times 5$ in 1D. The current deposition substep likewise sees a speedup of about $\times 2$ in 2D, and about $\times 5$ in 1D. The cost for communications is also significantly reduced in 1D and 2D, as the meshblocks communicate less data with fewer neighboring blocks.

\section{Tests}
\label{sec:tests}

\entity~comes with a set of fairly general pre-made problem generators that can either be used as is, or modified further to meet the specific requirements of the user. Below, we use these problem generators to both demonstrate the benchmarks done to test the validity of the code and also discuss specific settings and boundary conditions that went into designing these tests. For brevity and readability, all the vector components we use below are in the orthonormal coordinate system, i.e., $E^x$, $B_y$, $J^r$ etc are all defined in the coordinates with basis vectors of $\hat{\bm{x}},\hat{\bm{y}},\hat{\bm{z}}$, or $\hat{\bm{r}},\hat{\bm{\theta}},\hat{\bm{\phi}}$.

\subsection{Cartesian coordinates}

Note that all the problem generators presented in this section are written in a general form and are thus available in all dimensions -- 1D/2D/3D.


\subsubsection{Streaming instabilities: two-stream (1D) and Weibel filamentation (2D)}
\label{sec:pgen_streaming}

Plasma beam instabilities possess a vast history of theoretical investigations \citep[e.g.,][]{Bohm_1949, Fried_1959, Bret_2010}, making them a perfect testbed for numerical codes, with the two-stream and Weibel instabilities being most often taken as a benchmark of PIC codes \citep[e.g.,][]{Bacchini_2023, Croonen_2024}. For the demonstrations in this section, we used the built-in \texttt{streaming} problem generator, which can initialize an arbitrary number of plasma species of given temperatures and densities distributed uniformly and streaming with respect to each other with given velocities.

\begin{figure*}
	\centering
	\includegraphics[width=0.8\textwidth]{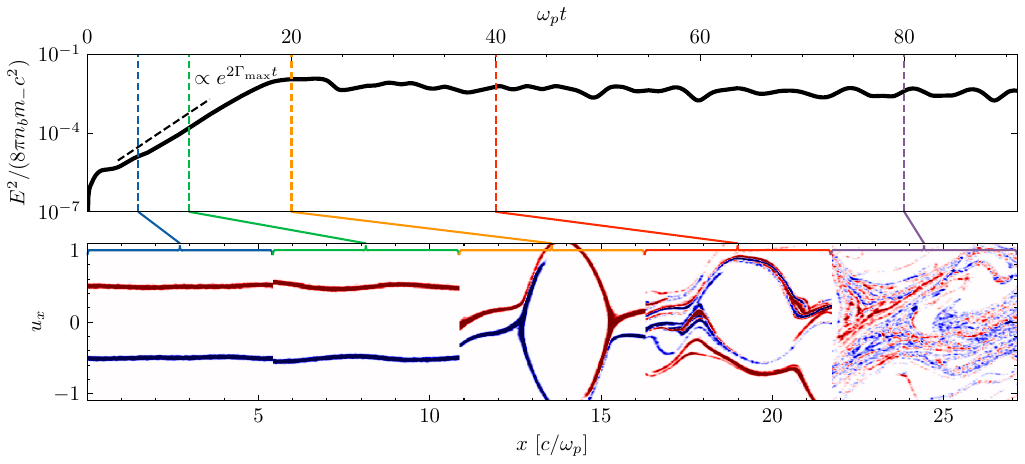}
	\caption{Two-stream instability test run results. Top: growth of the mean electric field energy density over time, normalized to the rest-mass energy density of particles. The black dashed line denotes the maximum growth rate of the instability $\Gamma_{\rm max} = \omega_{p}/(2\gamma_{b})^{3/2}$. Bottom: phase-space density of the electrons at different times indicated with colored lines. Blue and red colors are used to distinguish the two populations of the initially counter-streaming electrons.
	}
	\label{fig:twostream}
\end{figure*}

We start with the simple case of the relativistic two-stream instability. We study this instability by initializing two cold counter-propagating electron beams of densities $n_b/2$ in the neutralizing background of static ions, propagating in the $\pm x$-directions with the oppositely oriented velocities of $\pm u_{\rm b}$. The dispersion relation for such a system is then given by
\begin{equation}\label{eq:two_stream_dispersion}
	1 - \frac{\omega_p^2}{2\gamma_b^3}\left[\frac{1}{(\omega - kv_{b})^2} + \frac{1}{(\omega + k v_{b})^2}\right]=0.
\end{equation}
Here, the beam Lorentz factor $\gamma_{b} = \sqrt{1 + u_{b}^2}$, the beam three-velocity $v_{b} = u_{b}/\gamma_{b}$, and the plasma frequency $\omega_{p} = \sqrt{4\pi n_b q^2/m_-}$. The growth rate, $\Gamma = {\mathfrak{Im}}(\omega)$, of the two-stream instability is defined by the roots of the equation~\eqref{eq:two_stream_dispersion}. The maximum growth rate $\Gamma_{\rm max} = \omega_{ p}/(2\gamma_{ b})^{3/2}$ is achieved at $k_{\rm max} = \sqrt{3}\omega_{p}/\{(2\gamma_b)^{3/2} v_{b}\}$. For the test, we initialize the beams cold, $T_b = 10^{-4}m_-c^2$, and counter-streaming with velocities of $u_{b}=\pm0.5$. The box length is set to $L \approx 27 c/\omega_p = 12288\Delta x$, and the simulation lasts for about $100\omega_{p}^{-1}$. The results of the simulation are shown in Fig.~\ref{fig:twostream}. The evolution of the electric field energy is shown in the top row, demonstrating an excellent agreement with the predicted growth rate $\Gamma_{\rm max}$ (shown with a black dashed line) during the linear stage of the instability. The bottom row shows the evolution of the distribution of particles plotted versus the $x$ and $u_x$, showing the initial linear growth of the instability from numerical noise (we used a total of $64$ particles per simulation cell) and a subsequent non-linear saturation and mixing.

\begin{figure*}
	\centering
	\includegraphics[width=0.8\textwidth]{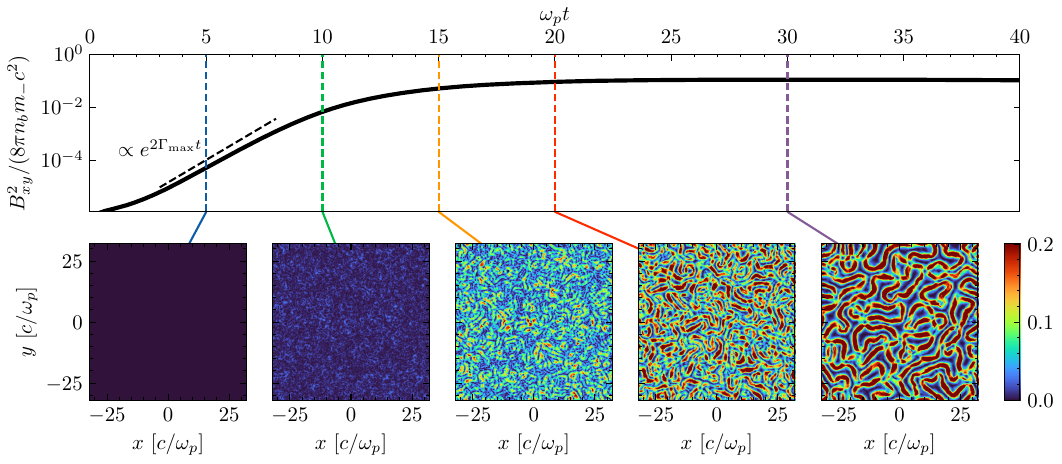}
	\caption{The results of the Weibel instability simulation. The top row shows the evolution of the total (in-plane) magnetic energy density as a fraction of the beam-plasma rest-mass energy density. Panels at the bottom show the amplitude of $|B_{xy}|$ with the same normalization at different stages of the simulation. Exponential growth with the maximum expected rate, $\Gamma_{\rm max}$, is shown with a dashed line.}
	\label{fig:weibel}
\end{figure*}

Weibel instability \citep{Fried_1959, Weibel_1959}, or current-filamentation instability, develops in the presence of significant temperature anisotropy, or in the presence of two counter-propagating flows. During the instability, an initially unmagnetized plasma can generate and amplify magnetic fields. To model this instability, we use the same setup as above, with the only difference that the beam velocities are now oriented along the $z$-direction, while the simulation itself is performed in the 2D $x\text{-}y$ plane. In such configuration, the in-plane component of the magnetic field, $B_{xy}\equiv\sqrt{B_x^2+B_y^2}$, will be generated and amplified with the linear growth rate $\Gamma_{\rm max} = v_{ b}\omega_{p}/\gamma_b^{1/2} $, achieved asymptotically at high wavenumbers $ck/\omega_p\gtrsim10$ \citep{Grassi_2017}. We increase the size of the simulation box to $L = 32~c/\omega_p$ to ensure that this asymptotic behavior is correctly captured in our simulation; as before, we evolve the system for $100\omega_{p}t$. The development of the instability leads to the formation of the filaments, see Fig.~\ref{fig:weibel}, where we plot the evolution of the in-plane magnetic field, $B_{xy}$. The filaments grow and merge with time. The magnetic field itself is significantly amplified, and the growth rate, again, shows excellent agreement with the theoretical maximum growth rate at linear stages of the evolution of the instability.\footnote{Minor discrepancies (within $5\dots 10\%$) of the growth rate are due to the fact that we do not initialize any perturbations in the box, meaning the modes start growing purely from the numerical noise.}



\subsubsection{Magnetic reconnection (2D)}

Relativistic magnetic reconnection has been extensively studied numerically in the past decade as a viable mechanism for extracting magnetic field energy in astrophysical systems, especially compact objects, and powering some of the most luminous -- transient and persistent -- observational phenomena \citep{Sironi.Spitkovsky_2014, Guo.etal_2014, Werner.etal_2016}. Typical kinetic simulations of the process initialize pair-plasmas in the so-called Harris equilibrium, with a cold, $T_\pm\ll m_-c^2$, highly magnetized, $\sigma\equiv B^2/4\pi\rho_\pm c^2\gg 1$, background, and a thin current layer with hot overdense drifting plasma supporting the switch of polarity of the magnetic field: from $+B\hat{\bm{x}}$ for $y>0$ to $-B\hat{\bm{x}}$ for $y < 0$. The reconnection process is then triggered by a small perturbation at the center of the layer, after which the entire layer starts reconnecting, forming relativistic outflows, magnetic islands (plasmoids), and x-points (where the actual dissipation takes place, $E>B$). To allow the initial transient that still carries the memory of the artificial Harris equilibrium to pass and focus on studying the steady state of the system, we ``open'' the boundaries in the $\pm \hat{\bm{x}}$ after roughly half of a light-crossing time of the domain along the layer (this is done by using the \texttt{ABSORB} boundaries for particles). Since the reconnecting layer advects plasma from upstream and pushes it out after reprocessing, we also provide a steady supply of plasma far upstream (close to the $\pm\hat{\bm{y}}$ boundaries), by measuring the density deficit in a thin layer along the external wall. Near the boundaries, we reset all the fields to their initial values in a thin buffer region to effectively absorb any incoming waves (by using the \entity's built-in \texttt{MATCH} boundaries), where we match the unmodified fields in the box to the target fields close to the boundary:

\begin{equation}
	F = s F + (1-s)F_{\rm target},~~~\text{where}~s=\tanh{\left(\frac{|\Delta x|}{\Delta s}\right)},
\end{equation}

\noindent where $F$ is one of the $E$ or $B$ field components, $\Delta x$ is the distance of the given component from the boundary, and $\Delta s$ is some pre-defined smoothing distance.
\begin{figure*}
	\centering
	\includegraphics[width=\textwidth]{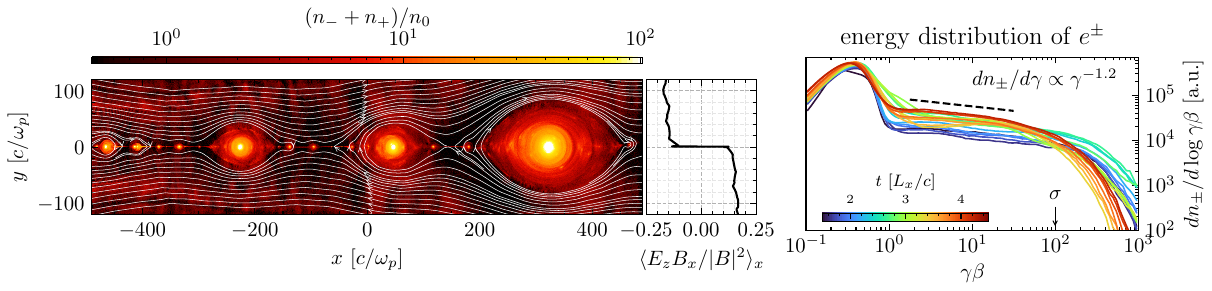}
	\caption{2D simulation of relativistic magnetic reconnection with cold highly-magnetized upstream, $\sigma=100$, at $t=4.5L_x/c$ ($L_x$ being the size of the domain in $x$). The main panel shows the snapshot of pair-plasma density together with the magnetic field lines, while the middle panel shows the average reconnection rate -- measured as $\langle E_z B_x/|B|^2\rangle_x$ -- as a function of $y$. The right panel shows the particle distribution function at different times.
	}
	\label{fig:reconnection}
\end{figure*}

The result of the 2D simulation after roughly $4.5$ light-crossing times is shown in Fig.~\ref{fig:reconnection}. The main panel shows the 2D distribution of the pair-plasma density measured in units of the upstream density, $n_0$. The middle panel shows the $x$-averaged reconnection rate (in the $y$-direction) as a function of $y$, while the rightmost panel shows the evolution of the particle distribution function. After the initial transient due to the strong perturbation in the middle of the layer, which lasts for about  $2...3~L_x/c$, a steady state is established with a hard power-law and an exponential cutoff near $\gamma\sim \sigma$, as is expected when $\sigma\gg 1$.



\subsubsection{Collisionless shock (2D)}
\label{sec:pgen-shock}

Collisionless shock waves shape the observational signatures of a broad range of astrophysical systems through various particle acceleration and radiative mechanisms \citep{Bell1978, Blandford1978, Drury1983, Blandford_1987, Treumann_2009, Bykov_2011, Sironi_2015, Marcowith_2016, Vanthieghem_2020, Levinson_2020}. The overarching challenge set by observations is to understand and model, from first principles, the nonthermal radiation they produce. Astrophysical shocks span a large range of scales, from kilometer-scale bow shocks around the Earth to megaparsec-scale virial shocks in galaxy clusters. Their signature are primarily characterized by the shock velocity $\beta_{\rm sh} \equiv v_{\rm sh}/c$ ($|\beta_{\rm sh}| \sim 10^{-4}$ for interplanetary shocks; $|\beta_{\rm sh}| \lesssim 1$ in active galactic nuclei), the plasma magnetization $\sigma \equiv B^2/4\pi w$,
where $w$ denotes the plasma enthalpy, ($\sigma \sim 10^{-9}$ for the forward shocks of gamma-ray burst afterglows; $\sigma \sim 10^{2}$ in shocks propagating in the inner magnetospheres of magnetars), as well as the upstream temperature and composition.

PIC simulations have been used extensively to study general properties of collisionless shocks in laboratory, space, and astrophysical environments, and characterize their acceleration efficiency \citep{Spitkovsky2008a, Amano_2009, Haugbolle_2011, Sironi2013, Guo_2014, Ryu_2019, Lemoine_2019, Lemoine_2019_III, Groselj_2022, Grassi_2023, Gupta_2024b, Groselj2024}; emission of coherent electromagnetic waves \citep{Hoshino_1992,Gallant_1994,Sironi_2021,Iwamoto_2017,Iwamoto_2018,Plotnikov_2018,Vanthieghem_2025}; thermalization and energy partition between species and the electromagnetic field \citep{Kato_2008, Kato_2010, Guo_2017, Guo_2018, Lemoine_2019_II, Bohdan2021, Vanthieghem_2022, Vanthieghem_2024, Gupta_2024a}. These essentially revolve around modeling the nonlinear development of plasma instabilities and the transport of particles in electromagnetic kinetic structures within the shock transition. In the following, we describe the general and standard shock setup implemented in \entity.

To initialize the shock, we start with a relativistically cold plasma, filling an elongated box ($L_x\gg L_y$) uniformly up to a certain distance from the left boundary: $0\le x <x_f$. We also push the plasma towards the $-x$ direction with a drift four-velocity, $u_x$, which ultimately determines the sonic Mach-number of the shock. The box is also filled with a magnetic field of strength $B_s$, at a certain angle $\theta$ w.r.t. the $x$ axis, and angle $\phi$ w.r.t. the $y\text{-}z$ plane:
\begin{equation}B_x=B_s\cos{\theta},~B_y=B_s\sin{\theta}\sin{\phi},~B_z=B_s\sin{\theta}\cos{\phi}.
\end{equation}

\noindent To accommodate the drift of the plasma, we initialize an electric field:
\begin{equation}
	E_x=0,~E_y=-\beta_xB_z,~E_z=\beta_xB_y,
\end{equation}

\noindent where $\beta_x\equiv u_x/\sqrt{1+u_x^2}$. The plasma is driven onto the left wall at $x=0$, where we enable the built-in \texttt{REFLECT} boundary conditions for the particles, which simply reflect the particle's position and the $x$-velocity component as they touch the wall. For the fields, we use the perfectly conducting boundary conditions (built-in \texttt{CONDUCTOR}), which force the following field components (which are defined at the exact boundary):
\begin{equation*}
	\begin{split}
		 & E^y_{\texttt{0,j+1/2,k}}=E^z_{\texttt{0,j,k+1/2}}=0, \\
		 & B^x_{\texttt{0,j+1/2,k+1/2}}=0.
	\end{split}
\end{equation*}

\noindent During the filtering of current densities, we assume that $J_{\texttt{-1/2,j,k}}^x=J_{\texttt{+1/2,j,k}}^x$, while $J_{\texttt{0,j+1/2,k}}^y=J_{\texttt{0,j,k+1/2}}^z=0$.

Additionally, to ensure the field components are also properly defined in the ghost zone beyond $x<0$ for later interpolation on the particle position, the \texttt{CONDUCTOR} boundaries enforce the following components:
\begin{equation}
	\begin{split}
		E^x_{\texttt{-di-1/2,j,k}}       & =E^x_{\texttt{di+1/2,j,k}},      \\
		E^y_{\texttt{-di-1,j+1/2,k}}     & =-E^y_{\texttt{di,j+1/2,k}},     \\
		E^z_{\texttt{-di-1,j,k+1/2}}     & =-E^z_{\texttt{di,j,k+1/2}},     \\
		B^x_{\texttt{-di-1,j+1/2,k+1/2}} & =-B^x_{\texttt{di,j+1/2,k+1/2}}, \\
		B^y_{\texttt{-di-1/2,j,k+1/2}}   & =B^y_{\texttt{di+1/2,j,k+1/2}},  \\
		B^z_{\texttt{-di-1/2,j+1/2,k}}   & =B^z_{\texttt{di+1/2,j+1/2,k}},  \\
	\end{split}
\end{equation}
\noindent where the value of \texttt{di} changes from $0$ to \texttt{N\_GHOSTS}. In the $+x$ direction, we have a moving injector replenishes the upstream plasma and resets the particle velocities in the region between $x_f(t) - c\beta_i \Delta t_i \le x<x_f(t)$, where $\beta_i$ is the speed of the injector, $\Delta t_i$, is the injection interval, while $x_f(t)$ is the current position of the injector. Boundary conditions in the $y$ or $z$ direction are periodic both for the fields and the particles.


\begin{figure*}
	\centering
	\includegraphics[width=0.8\textwidth]{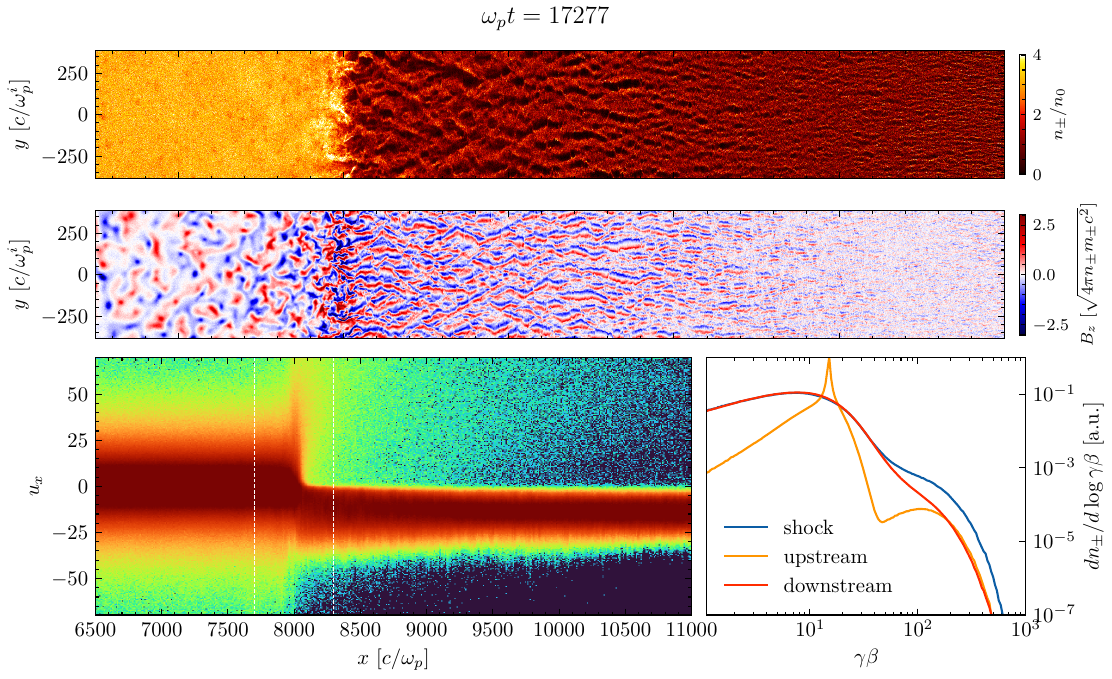}
	\caption{2D simulation of a collisionless weakly magnetized shock at a late stage. The top two panels show the plasma density and the out-of-plane component of the magnetic field, $B_z$, normalized to $\sqrt{4\pi n_\pm m_\pm c^2}$. The bottom panel on the left shows the phase-space distribution of both the electrons and the positrons. The right panel shows the distribution functions for both species computed in three different regions: near the shock (dashed lines in phase-space plot), upstream, and downstream.}
	\label{fig:shock}
\end{figure*}

In Figure~\ref{fig:shock}, we present the profile of a standard relativistic unmagnetized collisionless shock in pair-plasma. We initialize an unshocked cold plasma, $T_{\pm}\,=\,10^{-4}\, m_\pm c^2$, drifting with relativistic negative four velocity $u_x=-15$. The plasma is reflected from the left boundary, which implements the \texttt{REFLECT} boundary condition. The simulation frame thus corresponds to the downstream frame of the shock wave. The box is initially filled to $10\%$ of its width, $L_x\approx 20000\,c/\omega_p$, while the injector moves from its initial position with a speed of $\beta_{\rm inj}=1$ to the right. Here, $\omega_p\equiv \left(4\pi n_\pm e^2/m_\pm\right)^{1/2}$ which we resolve with $2.6$ cells. We also adopt the CFL number of $0.67$, meaning that $c\Delta t \approx 0.47 \Delta x$. Note, that to mitigate the Cherenkov emission upstream of the shock, we employ $32$ passes of a digital filter, and a modified stencil in Faraday's law (the details will be covered in paper III) with the following coefficients: $\beta_{xy}=\beta_{yx}=0.128$, and $\delta_x=-0.0005$, with the notations adopted from \cite{Blinne.etal_2018}.

Figure~\ref{fig:shock} corresponds to a snapshot at physical time $\omega_{p} t =17277$. We recover the characteristic filamented profile of the shock precursor, corresponding to $x\gtrsim 8000~c/\omega_p$, through the development and nonlinear evolution of the Weibel or current filamentation instability~\citep{Moiseev_1963,Medvedev_1999,Lyubarsky_2006,Achterberg_2007,Achterberg2007b,Ruyer_2016,Vanthieghem_2018,Ruyer_2018,Jikei_2024}. In the immediate shock downstream, the plasma is thermalized at the temperature $T_\pm\approx 7 m_{\rm e} c^2$ in good agreement with the relativistic jump conditions $T_\pm/m_\pm c^2=( \sqrt{1 + u_0^2} - 1 )/2$ in 2D. At shock crossing, about $1\%$ in number and $10\%$ in energy of the incoming particles are reflected in the upstream, escaping thermalization. These particles are injected into the distribution of nonthermal particles -- clearly visible on the phase-space and the energy distribution panels -- which are fed by Fermi acceleration through successive shock crossings. Finally, we recover the progressive scale-dependent collisionless damping of the downstream microturbulence \citep{Chang_2008, Lemoine_2015, Vanthieghem_2020}.


\subsubsection{Driven turbulence (3D)}
Turbulence is believed to be a ubiquitous phenomenon, characterizing plasma behavior in many astrophysical environments \citep[e.g.,][]{Zweibel1997, Balbus_1998, Elmegreen_2004, Mohapatra_2019}. PIC simulations became an essential tool to self-consistently model collisionless plasma turbulence phenomena, such as non-thermal particle acceleration \citep[e.g.,][]{Zhdankin_2017,Wong_2020,Vega_2022,Comisso_2024apjl,gorbunov_2025}, effects of non-ideal fields \citep[e.g.,][]{Comisso_2019}, various instabilities \citep[e.g.,][]{Davelaar_2020,Bacchini_2024}, radiation \citep[e.g.,][]{Zhdankin_2021,Groselj_2024,Nattila2024}, and plasma heating \citep[e.g.,][]{Zhdankin_2019prl,Zhdankin_2021_heating,Gorbunov_2025apjl}. \entity~offers a driven turbulence setup in a periodic box, allowing a framework to tailor to a specific problem of interest. Below, we outline the main ingredients required to achieve the steady-state turbulence in a PIC simulation, and implemented in \entity: a large-scale energy injection mechanism and a particle escape.

To inject energy at large scales, we ``stir'' the plasma in the box by introducing a Langevin antenna \citep{tenbarge2014}, which is used to drive an external current, $\bm{J}_{\rm ext}(t)$, added as a source term to the right-hand-side of Amp\'ere's equation,
\begin{equation}
	\frac{\partial \bm{E}}{c\partial t} = \nabla\times\bm{B} - \frac{4\pi}{c}(\bm{J}+\bm{J}_{\rm ext}),
\end{equation}
\noindent which is done by introducing a special class instance with a name \texttt{ext\_current} in the problem generator, which defines the individual current components. The external current is defined through a vector potential $\bm{A} = \{0,0, A_\parallel\}$, where the $z$-component, $A_{\parallel}$, is composed of multiple plane waves:
\begin{equation}
	A_\parallel (\bm{x},t) = \sum_{n}a_n(t) e^{i\bm{k}_n\cdot\bm{x}},
\end{equation}

\noindent where $a_n(t)$ are the complex amplitudes of the antenna, and $\bm{k}_n = \{k_{n}^x, k_{n}^y, k_{n}^z\}$, are the wave vectors of each of the modes. Considering $N$ total wave-vectors, the current is then computed as $\bm{J}_{\rm ext} = \nabla\times(\nabla\times{\bm{A}})$, which yields the following expression:
\begin{align}
	 & j_n(\bm{k}_n,\bm{x}) = \mathfrak{Re}(a_n) \cos{ (\bm{k}_n\cdot\bm{x})} - \mathfrak{Im}(a_n)\sin{(\bm{k}_n\cdot\bm{x})}, \nonumber \\
	 & J_{\rm ext}^x = -2\sum_{n=0}^{N/2} k_{n}^xk_{n}^z j_n(\bm{k}_n,\bm{x}),\nonumber                                                  \\
	 & J_{\rm ext}^y = -2\sum_{n=0}^{N/2} k_{n}^yk_{n}^z j_n(\bm{k}_n,\bm{x}),\nonumber                                                  \\
	 & J_{\rm ext}^z = 2\sum_{n=0}^{N/2} \left((k_{n}^{x})^2 + (k_{n}^{y})^2\right) j_n(\bm{k}_n,\bm{x}).
	\label{eq:antenna_current}
\end{align}

\noindent To obtain Eq.~\eqref{eq:antenna_current}, we have used the reality condition, $f(\bm{k}) = f^*(-\bm{k})$, where the $f^*$ is the complex conjugate of $f$. Given the size, $L$, of the cubic simulation domain, we drive the largest wavenumbers $\bm{k}_n L/(2\pi) = \{1,0,\pm 1\}$, $\{0,1,\pm1\}$, $\{-1,0,\pm 1\}$, $\{0,-1,\pm 1\}$ in the 3D case, and $\bm{k}_n L/(2\pi) = \{\pm1,0\}$, $\{0,\pm1\}$, $\{\pm1,\pm1\}$, $\{\mp1,\pm 1\}$, $\{\pm1,\mp1\}$ in 2D. The complex amplitudes, $a_n(t)$, are evolved according to the oscillating Langevin antenna equation \citep{tenbarge2014},

\begin{equation}
	\begin{split}
		a_n(t+\Delta t) = & ~ a_n(t)e^{-(\gamma_0 + i\omega_0)\Delta t}                         \\
		                  & +\sqrt{\frac{12\gamma_0}{\Delta t}}(u_{\rm re}+iu_{\rm im})\Delta t
	\end{split}
\end{equation}

\noindent where $\omega_0$ is the antenna frequency, $\gamma_0 > 0$ is the antenna decorrelation frequency, and $u_{\rm re},~u_{\rm im} \in[-1/2,1/2]$ are uniformly distributed random numbers. For the 2D simulations, we also evolve amplitudes with negative frequencies, $-\omega_0$, in order to ensure the presence of the counter-propagating waves. The initial amplitude $a_n(t=0)$ is defined to match the user-provided root-mean-square fluctuation of the magnetic field $\delta B$ as $|a_n(t=0)| = \delta B/\left(\sqrt{N}\left({(k_{n}^x)^2+(k_{n}^y)^2} \right)^{1/2}\right)$. We also initialize magnetic field perturbation to balance the initial external current which helps to avoid sharp transients during the initial stirring phase: $\left(\nabla\times\bm{B}\right)\bigr\rvert_{t=0}=(4\pi/c) \bm{J}_{\rm ext}(t=0)$.

The energy injected by the antenna in the absence of the explicit energy sink stops the system from reaching the steady-state. This results in a constant growth of energy in the simulations, and slow drift of particle energy distribution functions to higher energies. In \entity, we implement the mechanism of diffusive particle ``escape'' as explained in \citep{gorbunov_2025}. We track each particle's displacements  $\Delta l_{x,y}$ along $x,y$ axes from it's initial position; if the particle has diffused over some arbitrary (user-defined) \textit{escape} distance $l_{\rm esc} \le \max(\Delta l_x, \Delta l_y)$, we reset resample it's velocity from the initial Maxwellian distribution. In such a way, we mimic particles' diffusive escape, and provide a mechanism to cool down the most energetic particles in the system \citep{gorbunov_2025}.

We have conducted two runs: a three-dimensional one with the box size of $L=500 ~d_e$ in units of electron skin depth, $d_e\equiv c/\omega_p$, and a large-scale 2D run with $L=10^4~d_e$. Apart from the box sizes, the parameters of both simulations are identical. We resolve skin depth with $d_e=2\Delta x$, and let the system to evolve for total of $t=7(L/v_A)$ Alfv\'en crossing times, with Alfv\'en speed defined via magnetization $\sigma$ as  $v_A=c\sqrt{(\sigma/(\sigma+1))}$. We set $\sigma=1$ in our simulations. The antenna frequency is chosen to be slightly off resonant from the lowest Alfv\'en frequency of the system, $\omega_{\rm ant} = 0.9\omega_A=0.9\cdot2\pi(v_A/L)$.

\begin{figure*}
	\centering
	\includegraphics[width=0.8\textwidth]{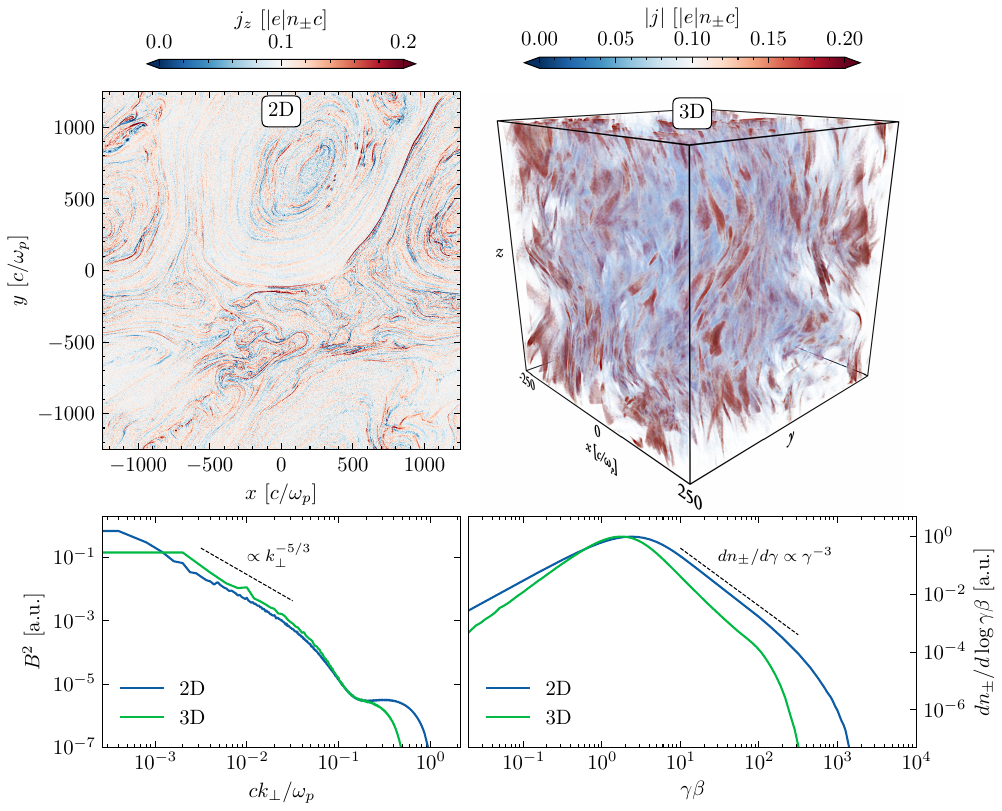}
	\caption{2D and 3D simulations of a driven turbulence. On the top left panel, we show the ($x$-$y$) box of the 2D run, depicting the $z$-component of the current density, while on the right, we show the volume rendering of the magnitude of the current $|\bm{j}|$ for the 3D run. Corresponding $k_\perp$-spectra of the magnetic field ($k_\perp\equiv (k_x^2+k_y^2)^{1/2}$) as well as the particle distributions for both runs are shown in the lower panels.}
	\label{fig:turb}
\end{figure*}

The results of the simulations are shown in Fig.~\ref{fig:turb}. In a steady-state, both systems develop thin turbulent current sheets (see top panels on Fig~\ref{fig:turb}). Both run produce turbulent magnetic fields, characterized by their spectra, shown in lower-left panel of Fig.~\ref{fig:turb}. We see that at large scales, both spectra produce a power law with $-5/3$ slope. At the kinetic scales $k_\perp\approx 1/d_e$, the spectra steepen as expected in this regime \cite{Zhdankin_2017}. Finally, our runs confirm that the turbulence can act as an effective source of nonthermal particles \cite[e.g.,][]{Zhdankin_2017, Comisso_2019, Zhdankin_2019prl, Gorbunov_2025apjl}, producing a power-law tail in particle distributions (Fig.~\ref{fig:turb}, bottom-right panel) at high energies for both runs.

\subsection{2.5D axisymmetric geometry}


In this Section, we discuss examples of applications of \entity~ for studying relativistic magnetospheres; in this example, we focus on neutron stars. The crust of a neutron star is a perfect conductor, thus on the surface of the star, which in the simulation domain is exactly at $r=r_{\rm in}$ (the inner radial boundary), we enforce the radial component of the magnetic field, $B_r$, and the tangential components of the electric field, $E_\theta$, $E_\phi$. To ensure that charges are completely perfectly absorbed by the neutron star surface, we also have a thin buffer zone of at least the size of the current filtering stencil, to which we extend the enforcement of these analytic field components. All of this is done via the built-in \texttt{ATMOSPHERE} boundary conditions for the fields. On the far end of the box, $r=r_{\rm out}$, we use the same  \texttt{MATCH} boundary conditions for the fields, and \texttt{ABSORB} conditions for the particles discussed above.

To model the extraction of charged particles from the neutron star atmosphere, we inject thermal pair-plasma just above the inner boundary using the built-in \texttt{ATMOSPHERE} boundary conditions for the particles (this also deletes the particles which cross the $r=r_{\rm in}$ boundary). The thermal plasma has a Boltzmann spatial distribution $n_\pm = n_a \exp{(-m_\pm\phi_g/T_\pm)}$, where, $\phi_g$ is a gravitational potential, and the peak density $n_a \approx 10 {n}_{\rm target}$. The ${n}_{\rm target}$ can either be the characteristic plasma number density in the magnetosphere (e.g., the Goldreich-Julian density for uniformly rotating magnetospheres), or the critical density required to conduct the current in twisted magnetospheres. The Boltzmann distribution is supported by adding a radial gravitational force to the particles' equations of motion, which is gradually turned off at a certain distance $\sim 2r_{\rm in}$. At each time step, the injector replenishes the escaping particles within one scale height from the stellar surface to maintain the Boltzmann distribution. The temperature, $T_\pm$, is selected to be low to prevent the quasi-neutral thermal outflow from the atmosphere from carrying the magnetospheric current, instead relying purely on the acceleration by the unscreened electric field. Below, we discuss the charge-separated flows in the monopolar and dipolar surface field geometries.

In this section, $r_{\rm in}$ corresponds to $R_*$ -- the radius of the neutron star. To keep particles' gyroradii close to zero, we also use the hybrid pusher which switches between the conventional algorithm by \cite{Boris_1970} and the guiding center approximation described by \cite{Bacchini.etal_2020} depending on the ratio of $|\bm{E}|/|\bm{B}|$ and the particle's nominal gyroradius at a given timestep, $\rho_L\approx \gamma\beta m c^2/q|\bm{B}|$. In the guiding center regime, the particle's momentum perpendicular to the local magnetic field is set to zero, effectively reducing its motion to that of its guiding center.\footnote{This is done by simply setting the \texttt{pusher} parameter for the given species to \texttt{"Boris,GCA"}, and specifying the critical Larmor radius for the transition between the two regimes. Equally, GCA pusher can be combined with Vay pusher by setting \texttt{"Vay,GCA"}.}.

In the two cases described below, we initialize an empty magnetosphere with either a monopolar or a dipolar magnetic field and no electric field. To mimic rotation of the star, the \texttt{ATMOSPHERE} boundaries for the fields enforce the following electric field components:

\begin{equation}
	\begin{split}
		E_r      & = \frac{\Omega}{c} r B_\theta \sin{\theta}, \\
		E_\theta & = -\frac{\Omega}{c} r B_r \sin{\theta},
	\end{split}
\end{equation}

\noindent where $\Omega$ is the user-specified rotation frequency (with its vector pointing along the pole at $\theta=0$), while $B_r$ and $B_\theta$ are the initial magnetic field components. We further refer to the quantity $R_{LC}=c/\Omega$ as the light cylinder. In both cases, we use a resolution of $\mathtt{N_r}\times \mathtt{N_\theta}=2048\times 1024$, with a quasi-spherical grid ($\log$ in $r$ and uniform in $\theta$), and a box which spans from $R_*$ to $50 R_*$. For the atmosphere, the peak density is chosen to be $n_a=10n_{GJ}^*$, where $n_{GJ}^*=\Omega B_*/(2\pi c|e|)$ is the Goldreich-Julian number density close to the stellar surface where the magnetic field strength is $B_*$. The $n_{GJ}^*$ is resolved with $10$ particles per each cell, and we additionally use $4$ current filter passes. The scale-height for the atmosphere is chosen to be $10^{-2} R_*$.

\subsubsection{Michel monopole}

\begin{figure*}[htb]
	\centering
	\includegraphics[width=0.8\textwidth]{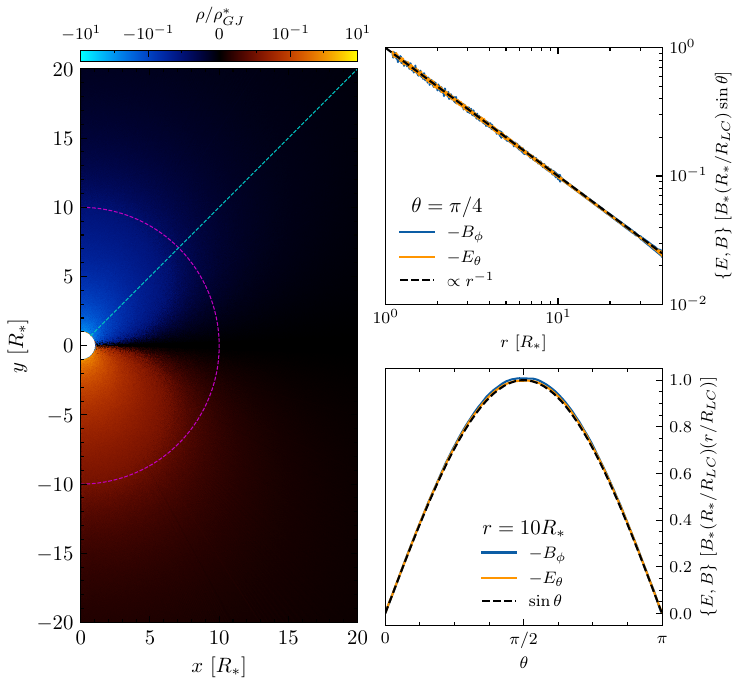}
	\caption{Solution for the rotating magnetic monopole, extracting the necessary plasma to satisfy the force-free condition from the surface atmosphere. On the left, we show the total charge density of the pair plasma $\rho$, and on the right, we show the dependencies of $B_\phi$ and $E_\theta$ on $r$ and $\theta$ together with the analytic predictions.}
	\label{fig:monopole}
\end{figure*}

In this test, we start with an initially radial magnetic field, $B_r = B_*(r/R_*)^{-2}$. After about a light-crossing time of the domain, a force-free steady state solution is established with $\bm{E}\cdot\bm{B}=0$ everywhere. This solution, shown in Fig.~\ref{fig:monopole}, has been described by \cite{Michel_1973}, and has the following form:

\begin{equation}
	\begin{split}
		B_r             & =B_* \left(\frac{r}{R_*}\right)^{-2},                               \\
		E_\theta=B_\phi & =-B_*\frac{R_*}{R_{LC}}\left(\frac{r}{R_*}\right)^{-1}\sin{\theta},
	\end{split}
\end{equation}

\noindent with the other field components being exactly zero. In the figure, we plot the charge density of the pair-plasma, $\rho\equiv e(n_+-n_-)$ normalized to the Goldreich-Julian density near the surface, $n_{GJ}^*$, as well as the $r$- and $\theta$-profiles of $B_\phi$ and $E_\theta$ together with the analytic expressions. The perfect agreement with the analytics means that the radial electric field at the surface is able to efficiently extract the necessary plasma from the atmosphere to satisfy the force-free, $\bm{E}\cdot \bm{B}=0$, condition throughout the domain.

\subsubsection{Disk-dome magnetosphere}

\begin{figure*}[htb]
	\centering
	\includegraphics[width=0.8\textwidth]{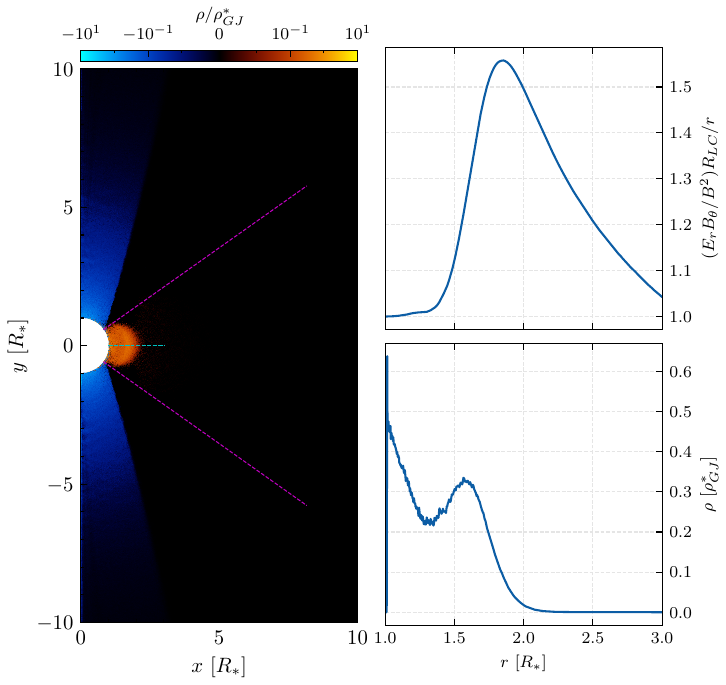}
	\caption{Solution with the dipolar magnetic field, where the surface charge extraction is unable to fill the entirety of the magnetosphere, resulting in the charge-separated disk-dome solution. Left panels show the charge density, while the right two panels show the radial profiles of the toroidal rotation velocity, $(\bm{E}\times\bm{B})_\phi$, and the charge density near the equator (shown with a cyan line).}
	\label{fig:diskdome}
\end{figure*}

In this test, instead of a monopolar magnetic field, we initialize a dipole, the magnetic axis of which is aligned with $\bm{\Omega}$. The resulting configuration after about a light-crossing time of the domain in $r$ is the so-called disk-dome solution \citep{Spitkovsky.Arons_2002, Petri.etal_2002}, with the two charged regions: the negatively charged dome, and the equatorial positively charged disk shown in Fig.~\ref{fig:diskdome}. The null line separating the two charge species is roughly at $\cos\theta=\pm 1/\sqrt{3}$ (shown with magenta lines). The disk of positrons, the radial density profile for which is shown in the same figure, rotates differentially, since not all field lines threading the disk are uniformly filled with plasma. In the upper right panel, we plot the rotation profile, $(E_rB_\theta/B^2)R_{LC}/r$, along the equator ($\theta=\pi/2$). Both of these profiles are in excellent agreement with both the theoretical predictions by \cite{Petri.etal_2002}, as well as previous 3D PIC simulations \citep[see, e.g., ][]{Philippov.Kramer_2022}.



\section{Summary}

In this paper, we introduced the first coordinate-agnostic GPU-accelerated open-source PIC code -- \entity. The code aims to accommodate the needs of a broad community of astrophysicists by providing a modular, customizable framework for a wide range of simulation setups and grid geometries, with no compromise on the performance and portability of the code. \entity~solves the Vlasov-Maxwell set of equations in general covariant form, which allows it to be easily extensible to more complex space-time geometries (e.g., see paper II for the general relativistic implementation). \entity~also comes pre-packaged with a powerful set of post-processing tools provided with the \texttt{nt2py} Python module. We have performed a set of scaling tests on most of the comprehensive GPU architectures currently available, with excellent performance across different major vendors and compatibility layers. In this paper, we also demonstrated a set of standard problem generators -- all included with the main source code -- which check the validity of the various algorithms in the code in different dimensions and grid geometries.

\begin{acknowledgments}
  The original inspiration behind \entity~came from numerous discussions  with and encouragement from Prof. William Dorland.

  Authors would also like to thank L. Sironi, A. Spitkovsky, D. Caprioli, C. Granier, M. Grehan, C. Thompson, and E. Quataert for their helpful comments and countless fruitful discussions. The developers would also like to acknowledge the technical support and advice provided by C. Trott (\texttt{Kokkos} team), V. Mewes (\texttt{Frontier} facility), R. Kakodkar (Princeton). The initial development of the code was supported by U.S. DOE grant DE-AC02-09CH11466, NSF Cyberinfrastructure for Sustained Scientific Innovation grant 2311800, and the NVIDIA Corporation Academic Hardware Grant Program. The authors would also like to acknowledge the OLCF Director's Discretion Project AST214 which enabled the testing of the code on the \texttt{Frontier} supercomputer. This work was facilitated by Simons Foundation (00001470, A.P.), and Multimessenger Plasma Physics Center (MPPC, A.P.), NSF grant No. PHY-2206607. A.C. is supported by Martin A. and Helen Chooljian Member Fund and the Fund for Memberships in Natural Sciences. L.M.B. is supported by NASA through grant 80NSSC24K0173 and NSF through grant AST-2510951. A.P. additionally acknowledges support by an Alfred P. Sloan Fellowship, and a Packard Foundation Fellowship in Science and Engineering. M.Z. acknowledges the support by NSF through grant PHY-2512037. The computations in this work were, in part, run at facilities supported by the Scientific Computing Core at the Flatiron Institute, a division of the Simons Foundation. The developers are pleased to acknowledge that the work was performed using the Princeton Research Computing resources at Princeton University which is a consortium of groups led by the Princeton Institute for Computational Science and Engineering (PICSciE) and Office of Information Technology's Research Computing. This research used resources of the Argonne Leadership Computing Facility, which is a U.S. DOE of Science User Facility operated under contract DE-AC02-06CH11357. We acknowledge LUMI-BE for awarding
  this project access to the LUMI supercomputer, owned by the EuroHPC Joint Undertaking, hosted by CSC (Finland) and the LUMI consortium through a LUMI-BE Regular Access call.

\end{acknowledgments}

\appendix


\section{Discretized Maxwell's equations in curvilinear coordinates}
\label{app:maxwell-discrete}

Below we present the full set of spatially discretized Maxwell's equations that are solved at substeps 2, 10, and 11 in the routine described in Section~\ref{sec:time-discretization}. The interpolation of fields to particle positions, the integration of particles' equations of motion, as well as the current deposition are done in exactly the same way as in conventional PIC algorithms, and we thus do not discuss these here. The discretized Faraday's law is then written in the following way:

\begin{align}
	\label{eq:discretized-faraday}
	\begin{aligned}
		 & \Delta_{\mathtt{n-1/2}}^{\mathtt{n+1/2}}
		\left[\AtNIJK{B^1}{*}{i}{\jPhalf}{\kPhalf}\right]=
		-\frac{\left\{
			\Delta_{\mathtt{j}}^{\mathtt{j+1}}\left[
				\CovAtNIJK{E}{3}{n}{i}{*}{\kPhalf}
				\right]-
			\Delta_{\mathtt{k}}^{\mathtt{k+1}}\left[
				\CovAtNIJK{E}{2}{n}{i}{\jPhalf}{*}
				\right]\right\}}{\sqrt{h_{(\mathtt{i,\jPhalf,\kPhalf})}}}
		,                                           \\
		 & \Delta_{\mathtt{n-1/2}}^{\mathtt{n+1/2}}
		\left[\AtNIJK{B^2}{*}{\iPhalf}{j}{\kPhalf}\right]=
		-\frac{\left\{
			-\Delta_{\mathtt{i}}^{\mathtt{i+1}}\left[
				\CovAtNIJK{E}{3}{n}{*}{j}{\kPhalf}
				\right]+
			\Delta_{\mathtt{k}}^{\mathtt{k+1}}\left[
				\CovAtNIJK{E}{1}{n}{\iPhalf}{j}{*}
				\right]\right\}}{\sqrt{h_{(\mathtt{\iPhalf,j,\kPhalf})}}}
		,                                           \\
		 & \Delta_{\mathtt{n-1/2}}^{\mathtt{n+1/2}}
		\left[\AtNIJK{B^3}{*}{\iPhalf}{\jPhalf}{k}\right]=
		-\frac{\left\{
			\Delta_{\mathtt{i}}^{\mathtt{i+1}}\left[
				\CovAtNIJK{E}{2}{n}{*}{\jPhalf}{k}
				\right]-
			\Delta_{\mathtt{j}}^{\mathtt{j+1}}\left[
				\CovAtNIJK{E}{1}{n}{\iPhalf}{*}{k}
				\right]\right\}}{\sqrt{h_{(\mathtt{\iPhalf,\jPhalf,k})}}}
		;
	\end{aligned}
\end{align}

\noindent while the Amp\'ere's law with the conformal currents $\mathcal{J}^i$ is written as:

\begin{align}
	\label{eq:discretized-ampere}
	\begin{aligned}
		 & \Delta_{\mathtt{n}}^{\mathtt{n+1}}
		\left[\AtNIJK{E^1}{*}{\iPhalf}{j}{k}\right]=
		\frac{\left\{
			\Delta_{\mathtt{j-1/2}}^{\mathtt{j+1/2}}\left[
				\CovAtNIJK{B}{3}{n+1/2}{\iPhalf}{*}{k}
				\right]-
			\Delta_{\mathtt{k-1/2}}^{\mathtt{k+1/2}}\left[
				\CovAtNIJK{B}{2}{n+1/2}{\iPhalf}{j}{*}
				\right]\right\}
			- 4 \pi \AtNIJK{\mathcal{J}^1}{n+1/2}{\iPhalf}{j}{k}
		}{\sqrt{h_{(\mathtt{\iPhalf,j,k})}}}
		,                                     \\
		 & \Delta_{\mathtt{n}}^{\mathtt{n+1}}
		\left[\AtNIJK{E^2}{*}{i}{\jPhalf}{k}\right]=
		\frac{\left\{
			-\Delta_{\mathtt{i-1/2}}^{\mathtt{i+1/2}}\left[
				\CovAtNIJK{B}{3}{n+1/2}{*}{\jPhalf}{k}
				\right]+
			\Delta_{\mathtt{k-1/2}}^{\mathtt{k+1/2}}\left[
				\CovAtNIJK{B}{1}{n+1/2}{i}{\jPhalf}{*}
				\right]\right\}
			- 4 \pi \AtNIJK{\mathcal{J}^2}{n+1/2}{i}{\jPhalf}{k}
		}{\sqrt{h_{(\mathtt{i,\jPhalf,k})}}}
		,                                     \\
		 & \Delta_{\mathtt{n}}^{\mathtt{n+1}}
		\left[\AtNIJK{E^3}{*}{i}{j}{\kPhalf}\right]=
		\frac{\left\{
			\Delta_{\mathtt{i-1/2}}^{\mathtt{i+1/2}}\left[
				\CovAtNIJK{B}{2}{n+1/2}{*}{j}{\kPhalf}
				\right]-
			\Delta_{\mathtt{j-1/2}}^{\mathtt{j+1/2}}\left[
				\CovAtNIJK{B}{1}{n+1/2}{i}{*}{\kPhalf}
				\right]\right\}
			- 4 \pi \AtNIJK{\mathcal{J}^3}{n+1/2}{i}{j}{\kPhalf}
		}{\sqrt{h_{(\mathtt{i,j,\kPhalf})}}}.
	\end{aligned}
\end{align}

\noindent In the equations above, we used the following conventions for brevity:

\begin{equation*}
	\begin{aligned}
		 & \Delta_{\mathtt{n}}^{\mathtt{n+1}}\left[{}^{(\mathtt{*})}F\right] \equiv \frac{{}^{(\mathtt{n+1})}F-{}^{(\mathtt{n})}F}{\Delta t};         \\
		 & \Delta_{\mathtt{i}}^{\mathtt{i+1}}\left[F_{(\mathtt{*,j,k})}\right] \equiv \frac{F_{(\mathtt{i+1,j,k})}-F_{(\mathtt{i,j,k})}}{\Delta x^1}, \\
		 & \Delta_{\mathtt{j}}^{\mathtt{j+1}}\left[F_{(\mathtt{i,*,k})}\right] \equiv \frac{F_{(\mathtt{i,j+1,k})}-F_{(\mathtt{i,j,k})}}{\Delta x^2}, \\
		 & \Delta_{\mathtt{k}}^{\mathtt{k+1}}\left[F_{(\mathtt{i,j,*})}\right] \equiv \frac{F_{(\mathtt{i,j,k+1})}-F_{(\mathtt{i,j,k})}}{\Delta x^3};
	\end{aligned}
\end{equation*}

\noindent where
\begin{equation*}
	\begin{aligned}
		\left.\begin{array}{lr}
			      F_{i(\mathtt{i,j,k})} \equiv h_{ii(\mathtt{i,j,k})}F^i_{(\mathtt{i,j,k})} \\
			      h_{ii(\mathtt{i,j,k})} \equiv h_{ii}\left(x^i_{(\mathtt{i,j,k})}\right)   \\
			      h_{(\mathtt{i,j,k})} \equiv h\left(x^i_{(\mathtt{i,j,k})}\right)          \\
		      \end{array}\right\} \text{for}~i=\{1,2,3\}.
	\end{aligned}
\end{equation*}

\noindent Note, that the currents, $\mathcal{J}^i$ in \eqref{eq:discretized-ampere} are the conformal currents, and are related to the physical contravariant currents as $\mathcal{J}^i\equiv \sqrt{h}J^i$.


\section{Quasi-spherical coordinate transformation}
\label{app:qspherical}

In Section~\ref{sec:spatial-discretization}, we discussed the quasi-spherical coordinate grid, which stretches the sizes of cells both in $r$ and $\theta$ directions. In particular, we introduced intermediate uniformly discretized coordinates, $\mathcal{R}$, and $\mathcal{T}$, defined through the transformations $r=r_0+e^\mathcal{R}$, and $\theta =\mathcal{T}+2\vartheta \mathcal{T}(1-2\mathcal{T}/\pi)(1-\mathcal{T}/\pi)$, with $r_0$ and $\vartheta$ parameterizing the transformation. As the particle pusher in quasi-spherical coordinates updates the positions in global Cartesian coordinates, it is thus necessary to also define a backwards transformation $(\mathcal{R},\mathcal{T})\rightarrow(r,\theta)$. For the radial component, the transformation is trivial: $\mathcal{R}=\log{(r-r_0)}$, however to reconstruct $\mathcal{T}$, it is required to find a root for a cubical polynomial (this is only performed when $\vartheta\ne 0$). For completeness, below we present the correct solution to the equation which recovers the value of $\mathcal{T}$ given the $\theta$-coordinate and the parameter $\vartheta$:

\begin{equation}
	\label{eq:qsph-theta-transform}
	\begin{split}
		\mathcal{T} & = \frac{\pi^{2/3}}{12}\left(6 \pi^{1/3} - 2^{4/3}(3\pi)^{2/3}\frac{1-\vartheta}{\Theta} + 2^{2/3} 3^{1/3} \frac{\Theta}{\vartheta}\right), \\
		            & \text{where}~~~\Theta\equiv \left(
		\sqrt{3\vartheta^3} \left\{
		108\vartheta\theta^2-108\vartheta\pi\theta+\pi^2(4-\vartheta)(1+2\vartheta)
		\right\}^{1/2}
		- 9 \vartheta^2 (\pi - 2 \theta)
		\right)^{1/3}.
	\end{split}
\end{equation}

\bibliography{references}{}
\bibliographystyle{aasjournalv7}

\end{document}